\documentclass[aps,prl,twocolumn,superscriptaddress]{revtex4-2}

\usepackage{amsmath,amssymb,amsfonts,color,graphicx,tabularx,dsfont,braket,tikz}

\usepackage[scr=rsfs]{mathalpha}

\usepackage[unicode=true,colorlinks=true]{hyperref}

\hypersetup{linkcolor=blue,citecolor=blue,urlcolor=blue}

\renewcommand{\ol}[1]{\overline{#1}}
\newcommand{\comments}[1]{}
\newcommand{\mb}[1]{\mathbf{#1}}
\renewcommand{\cal}[1]{\mathcal{#1}}

\def\id{\mathds{1}}

\def\Z{\mathbb{Z}}

\begin{document}

\title{Impurity screening by defects in (1+1)$d$ quantum critical systems}

\author{Ying-Hai Wu}
\thanks{These two authors contributed equally.}
\affiliation{School of Physics and Wuhan National High Magnetic Field Center, Huazhong University of Science and Technology, Wuhan 430074, China}

\author{Yueshui Zhang}
\thanks{These two authors contributed equally.}
\affiliation{Faculty of Physics and Arnold Sommerfeld Center for Theoretical Physics, Ludwig-Maximilians-Universit\"at M\"unchen, 80333 Munich, Germany}

\author{Hong-Hao Tu}
\affiliation{Faculty of Physics and Arnold Sommerfeld Center for Theoretical Physics, Ludwig-Maximilians-Universit\"at M\"unchen, 80333 Munich, Germany}
\affiliation{Institut f\"ur Theoretische Physik, Technische Universit\"at Dresden, 01062 Dresden, Germany}

\author{Meng Cheng}
\affiliation{Department of Physics, Yale University, New Haven, CT 06511-8499, USA}

\begin{abstract}
We propose a novel mechanism of impurity screening in (1+1)$d$ quantum critical states described by conformal field theories (CFTs). An impurity can be screened if it has the same quantum numbers as some gapless degrees of freedom of the CFT. The common source of these degrees of freedom is the chiral primary fields of the CFT, but we uncover that topological defect lines of the CFT may also take this role. Theoretical analysis relies on the insight that the impurities can be interpreted as edge modes of certain symmetry-protected topological (SPT) states. By stacking a SPT state with a CFT, one or two interfaces on which the SPT edge modes reside are created. If screening occurs due to topological defect lines, a symmetry-enriched CFT with exotic boundary states are obtained. The boundary conditions that appear in these cases are difficult to achieve using previously known methods. As a concrete example, we consider a spin-1 chain whose bulk is described by the SU(3)$_{1}$ CFT and edges are coupled to spin-1/2 impurities. We demonstrate that both the low-energy eigenstates and the extracted Affleck-Ludwig entropy are in excellent agreement with our theoretical predictions.
\end{abstract}

\maketitle

{\em Introduction} --- Quantum many-body systems with strong interactions are intrinsically difficult to study. The physics in one spatial dimension is most well-understood thanks to the experiences accumulated during the past century. It is useful to divide phases of matter into gapless and gapped ones. A prominent example of 1D gapless phase is the spin-1/2 Heisenberg model, which was solved exactly by Bethe shortly after its inception~\cite{Bethe1931} and placed in the more general framework of conformal field theory (CFT) in the 1980s~\cite{Belavin1984}. The study of gapped phases is greatly facilitated by the presence of energy gaps. The past two decades witness tremendous progress toward a complete classification of gapped phases with or without symmetries~\cite{Schnyder2008,Kitaev2009,Fidkowski2011,ChenX2011,Schuch2011,Kitaev2011,LuYM2012,ChenX2013,GuZC2014-2,WangC2014,Kapustin2015,Kapustin2017,ChengM2018,WangQR2018,Barkeshli2019,Aasen2021,Barkeshli2022,Bulmash2022}. For a 1D bosonic system that preserves certain global symmetries, all gapped phases can be fully classified~\cite{ChenX2011, Schuch2011}. The nontrivial ones are called symmetry-protected topological (SPT) states. If a SPT state is realized with open boundary condition (OBC), there would be edge modes that transform as a projective representation of the symmetry group. 

An important open problem is to establish a similar classification for gapless states. Recent works have uncovered some quantum critical systems with topologically protected edge modes~\cite{Keselman2015,Scaffidi2017,Ruhman2017,FuruyaS2017,Verresen2018,Parker2018,JiangHC2018,YaoY2019,JiWJ2020,Verresen2021-2,Thorngren2021,HidakaY2022,YuXJ2022,LiLH2024}. One route towards such ``symmetry-enriched" gapless states is stacking well-known gapless states with gapped SPT states. If the system has OBC, an edge mode of the SPT state can be interpreted as an ``impurity" that is added to the edge of the gapless state. In spite of the coupling with the gapless bulk, the edge modes could survive in some cases and lead to ground-state degeneracy in finite-size energy spectra.

It is also possible for the impurity to be ``screened" by the gapless bulk. This behavior is familar in the Kondo effect, where magnetic impurity is screened or overscreened by conduction electrons, depending on the number of conduction channels and coupling strengths~\cite{Hewson-Book}. More generally, we need to have certain gapless degrees of freedom in the bulk to screen the impurity. For a rational CFT, its gapless modes can be described using chiral (and anti-chiral) primary fields. If these fields carry the same quantum numbers as the impurity, it is natural to expect that the impurity is screened. However, this picture is inadequate because other screening mechanisms can be found. In this Letter, we propose that an edge impurity may be screened by topological defect lines (TDLs) of CFTs~\cite{Chang:2018iay}. This claim is supported by general analytical results and numerical calculations in spin chain models.

{\em Generalities} --- The question of whether a distinct symmetry-enriched CFT is created, or equivalently, whether the impurity can be screened, could be formulated using boundary CFT (BCFT)~\cite{IshibashiN1989,Cardy1989,Affleck1991-2}. To be concrete, the main text only studies spin chains with global SO(3) spin rotation symmetry, which acts \emph{faithfully} on the low-energy degrees of freedom in the CFT \footnote{This excludes the scenario in which certain symmetries only act on the gapped sector~\cite{Verresen2021-2}.}. The SPT state stacked on the CFT should also respect this SO(3) symmetry, so it actually belongs to the Haldane SPT phase of spin-1 chain~\cite{Haldane1983-2}. Its edge mode, or the impurity spin, transforms as a half-integer spin arising from projective representations of SO(3).

If the impurity is screened at low energy, the CFT must acquire a new conformal boundary condition~\footnote{More precisely, the boundary condition must be simple, so the CFT on an interval has a unique ground state.}. For the Kondo problems, it is well-known that magnetic impurities act as boundary condition changing operators~\cite{Affleck1990,Affleck1991}. In BCFT, conformal boundary conditions can be understood most conveniently in terms of boundary states, which are special physical states in the CFT Hilbert space on a circle with short-range correlation. For each boundary state, the associated boundary condition can be viewed as the interface between one region described by the CFT and another region by the boundary state. If a boundary state $\ket{B}$ preserves SO(3), one may ask to which SPT phase it belongs. This picture suggests that a half-integer spin impurity would be screened by the CFT if and only if there exists a boundary state that belongs to the Haldane phase.

It is very difficult to fully classify the boundary states of a given CFT~\cite{Behrend1998,Behrend2000}. For rational CFTs with diagonal partition functions, a frequently used set of boundary conditions are described by the Cardy states~\cite{Cardy1989}, which are in one-to-one correspondence with the chiral primary fields. The universal properties of a Cardy state $\ket{a}$ are determined by the field $a$ labeling it. If $a$ carries half-integer (integer) spin representation under the global SO(3) symmetry, the corresponding Cardy state is in the Haldane (trivial) phase. Consequently, the impurity can be completely screened by the Cardy state $\ket{a}$ if $a$ carries half-integer spin. This agrees with our intuition that screening is done by the CFT scaling fields. For example, we consider the SU(2)$_{1}$ CFT realized in the spin-1/2 Heisenberg chain. If a spin-1/2 impurity is attached to one end, it can be screened by the Cardy state corresponding to the spin-1/2 chiral primary field. We define $\ket{B_0}$ as a reference boundary state that preserves the global symmetry, and it is a trivial state. 

However, the Cardy states do not exhaust all possible boundary conditions. One way to generate new boundary states from known ones is ``fusing" TDLs~\cite{Chang:2018iay,FukusumiY2021}. For our purpose, TDLs can be viewed as generalizations of global symmetry transformations that include non-invertible operations such as the Kramers-Wannier duality~\cite{Bhardwaj:2017xup,KongL2020,McGreevy:2022oyu,Cordova:2022ruw}. In fact, the Cardy states themselves can be obtained by fusing a special set of TDLs, namely the Verlinde lines, to the Cardy state $\ket{\id}$, where $\id$ represents the identity primary field. The screening mechanism discussed above can be generalized to TDLs. In particular, if the TDL is invariant under the SO(3) symmetry, it can also be associated with a SU(2) representation. Intuitively, this is the spin representation at the end point of an open TDL, i.e., a defect operator. More details about TDLs are given in the Supplemental Material (SM)~\cite{Append}. Let us consider the boundary state $\ket{\cal{D}}=\cal{D}\ket{\id}$ obtained by fusing the TDL $\cal{D}$ with $\ket{\id}$. If $\cal{D}$ is a simple TDL, $\cal{D}\ket{B_0}$ is a simple boundary state. On the contrary, $\ket{\cal D}$ belongs to the Haldane phase if $\cal{D}$ is associated with half-integer spin representations.

This defect screening mechanism shall be demonstrated using a critical spin-1 chain. We study two diagnostics that correspond to the ``open" and ``closed" channels of BCFT on lattice. For the open channel, energy spectra of finite-size systems are computed numerically, which confirm theoretical results about the partition functions. For the closed channel, the celebrated spin-1 Affleck-Kennedy-Lieb-Tasaki (AKLT) state~\cite{Affleck1987-1} is identified as an \emph{integrable} boundary state of the spin chain that realizes the CFT. The overlap between the AKLT state and spin chain ground state is computed using Bethe ansatz solution of the latter. Its dependence on the system size allows us to extract a key quantity of conformal boundary states, namely the Affleck-Ludwig (AL) entropy~\cite{Affleck1991-2}. Thanks to integrability, analytical result in the thermodynamic limit is obtained, which agrees perfectly with BCFT prediction.

\begin{figure}[ht]
\centering
\includegraphics[width=0.40\textwidth]{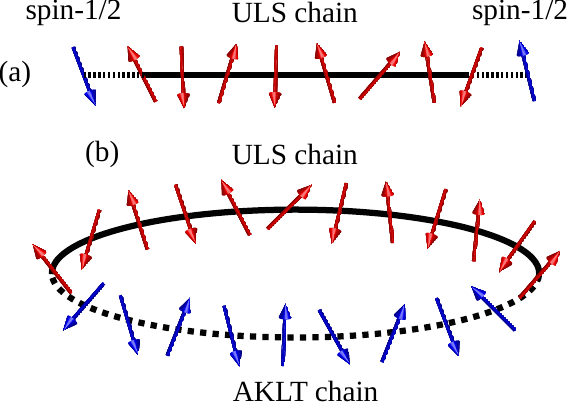}
\caption{Schematics of our model. (a) The middle part is a spin chain whose low-energy physics is described by a CFT. Its two edges are attached with spin-1/2 impurities. (b) The spin-1/2 impurities in panel (a) can be viewed as the edge states of the AKLT chain.}
\label{fig:stacking}
\end{figure}

{\em Spin-1 chain: SU(3)$_{1}$ CFT} --- The spin-1 lattice model with Hamiltonian
\begin{equation}
    H_{S=1}(\gamma) = \sum_{j} \left[ \mb{S}_{j} \cdot \mb{S}_{j+1} + \gamma\left( \mb{S}_{j} \cdot \mb{S}_{j+1} \right)^{2} \right]
\end{equation}
is employed to illustrate our theory. For $\gamma\in (-1,1)$, the ground state belongs to the Haldane phase and an open chain has one spin-1/2 edge mode on each end~\cite{Haldane1983-1,Haldane1983-2,Affleck1987-1,GuZC2009,Pollmann2012}. In particular, the $\gamma=1/3$ case is the AKLT model~\cite{Affleck1987-1}. The system enters a gapless phase when $\gamma>1$. The critical point at $\gamma=1$ is the Uimin-Lai-Sutherland (ULS) model, whose low-energy theory is the SU(3)$_{1}$ CFT~\cite{Uimin1970,LaiCK1974,Sutherland1975,Affleck1986-1,Affleck1988,ItoiC1997}. As illustrated in Fig.~\ref{fig:stacking} (a), we are interested in the ULS chain with one or both ends coupled to spin-1/2 impurities. Alternatively, the ULS chain may be connected to an AKLT chain whose edge modes are taken as impurities on the interfaces [see Fig.~\ref{fig:stacking} (b)]. In both cases, the SO(3) symmetry should be preserved by the couplings. 

If spin-1 is viewed as the fundamental representation of SU(3), the ULS model can be recast as the SU(3) Heisenberg chain (up to rescaling and overall shift). In other words, the symmetry of $H_{S=1}(\gamma=1)$ is enhanced from SO(3) to PSU(3). There are three primary fields in the SU(3)$_{1}$ CFT labeled by the SU(3) representations $\mb{0}$, $\mb{3}$, and $\bar{\mb{3}}$. The corresponding Cardy states are precisely the three SPT states with PSU(3) symmetry. Without loss of generality, the reference boundary state is chosen as $\ket{\mb{0}}$. If a spin-1/2 impurity is attached to one end of the ULS chain [which brings the symmetry down to SO(3)], it would appear that no Cardy state can screen the impurity: the $\mb{3}$ and $\bar{\mb{3}}$ representations simply reduce to spin-1 when the symmetry is broken down to SO(3). Nevertheless, the SU(3)$_{1}$ CFT occurs at a transition out of the Haldane phase, so the spin-1/2 edge modes could manifest themselves in the critical theory. In fact, orbifolding a $\Z_{2}$ symmetry $C$ in the SU(3)$_{1}$ CFT leads to the SU(2)$_{4}$ CFT. This symmetry is the charge conjugation of SU(3) that exchanges the two fundamental representations $\mb{3}$ and $\ol{\mb{3}}$~\footnote{More precisely, $C$ is the chiral $\Z_{2}$ symmetry that only acts in the left or the right sector. The charge conjugation acting on the full CFT is an invertible $\Z_{2}$ symmetry.}. After orbifolding, a twist operator becomes the spin-1/2 field of SU(2)$_{4}$. Therefore, the $C$ symmetry defect operator transforms as spin-1/2 under SO(3), so it can screen a spin-1/2 impurity to generate a new boundary condition. The associated TDL is denoted as $\cal{D}_{\frac{1}{2}}$.

Using the conformal embedding SU(2)$_{4}\subset$ SU(3)$_{1}$, the boundary condition and the partition function can be obtained explicitly~\cite{Append}. An essential insight is that SU(3)$_{1}$ is obtained from SU(2)$_{4}$ if its chiral algebra is extended using the SU(2) spin-2 chiral primary. In the larger chiral algebra, the spin-1 primary of SU(2)$_{4}$ splits into $\mb{3}$ and $\mb{\bar{3}}$ of SU(3)$_{1}$. Both spin-1/2 and 3/2 primaries turn out to be the same TDL $\cal{D}_{\frac{1}{2}}$. The fusion rule $\frac{1}{2}\times\frac{1}{2}=0 \; + \; 1$ of SU(2)$_{4}$ thus gives rise to 
\begin{eqnarray}
    \cal{D}_{\frac{1}{2}} \times \cal{D}_{\frac{1}{2}} = \mb{0} +  \mb{3} + \bar{\mb{3}}.
    \label{fusion-rule-D}
\end{eqnarray}
For a system with boundary conditions $A$ and $B$ at its two ends, the low-energy spectral information is fully encoded in the partition function $\mathcal{Z}_{A,B}$ of the CFT. After expanding $\mathcal{Z}_{A,B}$ as a power series $\sum_{j} n_{j} q^{h_{j}-\frac{c}{24}}$, it becomes clear that the eigenvalue $h_{j}$ is $n_{j}$-fold degenerate (up to overall normalization and shift). From the fusion rule \eqref{fusion-rule-D}, the system with boundary state $\ket{\cal{D}_{\frac{1}{2}}}$ on both ends has the partition function
\begin{eqnarray}
    \cal{Z}_{\cal{D}_{\frac{1}{2}},\cal{D}_{\frac{1}{2}}} &=& \chi_{\mb{0}} + \chi_{\mb{3}} + \chi_{\ol{\mb{3}}} \nonumber \\
    &=& q^{-\frac{1}{12}} \left( 1+6q^{\frac{1}{3}}+8q+\cdots \right).
\label{eq:PartFuncULS1}
\end{eqnarray}
If one end has boundary state $\ket{{\cal D}_{\frac{1}{2}}}$ and the other has $\ket{\mb{0}}$, the partition function is
\begin{equation}
    \cal{Z}_{\cal{D}_{\frac{1}{2}},\mb{0}} = \chi_{\frac{1}{2}} + \chi_{\frac{3}{2}} = q^{\frac{1}{24}} \left( 2+4q^{\frac{1}{2}}+6q+\cdots \right)
\label{eq:PartFuncULS3}
\end{equation}
due to the fact that $\mb{0}$ in SU(3)$_{1}$ is lifted back to $0 \; {\oplus} \; 2$ in SU(2)$_{4}$.

\begin{figure*}[ht]
\centering
\includegraphics[width=0.90\textwidth]{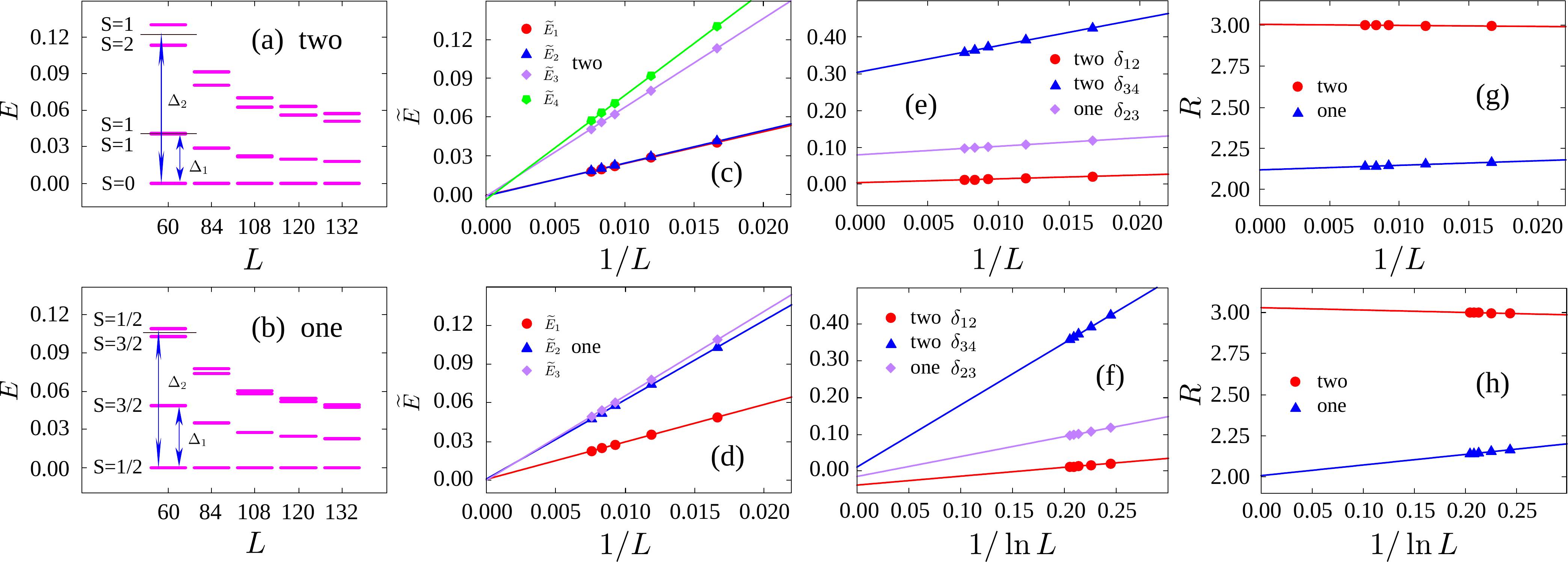}
\caption{Numerical results on the ULS chains with spin-1/2 impurities. (a,b) Energy spectra for the cases with two impurities or one impurity. The ground-state energy is shifted to zero. The total spin of each level is indicated. The averaged spacings $\Delta_{1,2}$ defined in the main text are illustrated. (c,d) Finite-size scaling analysis of the energy spacings $\widetilde{E}_{i} = E_{i}-E_{0}$. (e,f) Finite-size scaling analysis of the energy splittings $\delta_{12},\delta_{34}$ for two impurities and $\delta_{23}$ for one impurity. (g,h) Finite-size scaling analyis of the ratios $R=\Delta_{2}/\Delta_{1}$.}
\label{fig:ULS}
\end{figure*}

{\em Energy spectra} --- Both models in Fig.~\ref{fig:stacking} have been studied using the density matrix renormalization group (DMRG)~\cite{White1992}. Here we focus on the case with spin-1/2 impurities in Fig.~\ref{fig:stacking} (a) because larger systems can be accessed. Numerical results provide strong support for our theory, albeit with significant finite-size effects that will be analyzed carefully below. For the ULS-AKLT model in Fig.~\ref{fig:stacking} (b), the maximal available length is much smaller and the numerical results are less supportive~\cite{Append}. Each impurity is coupled to its nearest and next-nearest neighbors by Heisenberg terms~\cite{Append}. There are $L$ sites in the ULS part. It should be a multiple of $3$ such that the $\ket{\mb{0}}$ boundary state is realized when no impurity is attached. The total spin $\mb{S}^{2}=S(S+1)$ and its $z$-component $S_{z}$ are good quantum numbers, so it is sufficient to consider only the $S_{z}=0$ ($S_{z}=-\frac{1}{2}$) subspace for two impurities (one impurity). For an energy level with total spin $S$, we have $2S+1$ degenerate states. Energy spectra for multiple $L$ values are displayed in Fig.~\ref{fig:ULS} (a,b). The eigenvalues are named as $E_{0},E_{1},E_{2},\ldots$ in ascending order. When two impurities are attached, the lowest five levels have $S=0,1,1,2,1$. The uniqueness of ground state is consistent with complete screening of the edge modes. If there is only one impurity, the lowest four levels have $S=\frac{1}{2},\frac{3}{2},\frac{3}{2},\frac{1}{2}$. An inspection of the energy spacings $\widetilde{E}_{i} = E_{i}-E_{0}$ reveals that they decay linearly with $1/L$ [see Fig.~\ref{fig:ULS} (c,d)], as one would expect for CFTs.

It is apparent in Fig.~\ref{fig:ULS} (a) that $E_{1},E_{2}$ are almost degenerate whereas $E_{3},E_{4}$ are somewhat closer to each other than the rest ones. To make quantitative checks, we plot $\delta_{12}=(E_{2}-E_{1})/\widetilde{E}_{1}$ and $\delta_{34}=(E_{4}-E_{3})/\widetilde{E}_{1}$ versus $1/L$ or $1/\ln L$ in Fig.~\ref{fig:ULS} (e-f). Better fitting quality is found in the latter case. For example, the coefficent of determination for $\delta_{12}$ is $0.9956$ ($0.9997$) when $1/L$ ($1/\ln L$) is used. Logarithmic dependence is usually a signature of marginally irrelevant terms and their impacts diminish very slowly. We conclude that $E_{1},E_{2}$ collapse to a single level with higher degeneracy in the large $L$ limit and so do $E_{3},E_{4}$. This is reasonable since the bulk CFT has a larger symmetry group that is only broken by the boundary. It is thus established that the first excited level is $6$-fold degenerate ($E_{1},E_{2}$) and the second excited level is $8$-fold degenerate ($E_{3},E_{4}$), in full agreement with Eq.~\eqref{eq:PartFuncULS1}. For one impurity, $E_{1}$ is the first excited level with exact degeneracy $4$, whereas $E_{2}$ and $E_{3}$ collapse to a single level with total degeneracy $6$ as deduced from the plot of $\delta_{23}=(E_{3}-E_{2})/\widetilde{E}_{1}$ in Fig.~\ref{fig:ULS} (e,f). These results also agree with Eq.~\eqref{eq:PartFuncULS3}.

Based on these identifications, it is natural to define averaged spacings $\Delta_{1}=(E_{1}+E_{2})/2-E_{0},\Delta_{2}=(E_{3}+E_{4})/2-E_{0}$ for two impurities and $\Delta_{1}=E_{1}-E_{0},\Delta_{2}=(E_{2}+E_{3})/2-E_{0}$ for one impurity. According to BCFT predictions, the ratio $R=\Delta_{2}/\Delta_{1}$ should respectively approach $3$ and $2$ in the large $L$ limit for these two cases. For the system with $L=132$, we obtain $R=2.9994$ for two impurities and $R=2.1403$ for one impurity. It would appear that the former is a perfect match and obvious deviation of the latter should be attributed to finite-size effects. The situtation is actually more involved as one can see from the finite-size scaling of $R$ in Fig.~\ref{fig:ULS} (g,h). For one impurity, the fitting versus $1/\ln L$ has better quality and the extrapolated number is $2.0088$, which again suggests that marginally irrelevant terms are present. However, the fitting versus $1/\ln L$ is worse when there are two impurities and the extrapolated number is $3.0292$. We believe that the perfect match between theoretical and numerical values at $L=132$ is a coincidence. Indeed, finite-size effects are still important as reflected by the splitting between $E_{3}$ and $E_{4}$ that is clearly visible in Fig.~\ref{fig:ULS} (a). If the data points with $L<100$ are removed, the fitting gets better and the extrapolated number becomes $3.0188$. It is reasonable to speculate that $R$ would decrease slowly toward $3$ as $L$ increases.

{\em AL entropy from wavefunction overlap} --- Based on our analysis, the boundary state that screens a spin-1/2 impurity in the SU(3)$_{1}$ CFT should belong to the Haldane phase. In the lattice model, a natural candidate for this boundary state is the AKLT state at $\gamma=\frac{1}{3}$. We support this identification by showing that the AL entropy of the boundary state can be extracted from a unverisal term in the wavefunction overlap. For a CFT ground state $\ket{0}$ and a boundary state $\ket{B}$, one can define a $g$-factor $g_{B}=\braket{0|B}$ and the AL entropy is given by $\ln g_B$~\cite{Affleck1991-2}. To apply this result on lattice, the boundary state should be regularized such that~\cite{Calabrese:2006rx,Cardy:2017ufe}
\begin{equation}
    \braket{0|B_{\rm reg}} \cong g_{B} e^{-\alpha L},
\end{equation}
where $L$ is the system size and $\alpha$ is a non-universal constant (often referred to as the ``surface free energy density''). This formula offers a convenient method to extract $g$ from finite-size scaling of the wavefunction overlap~\cite{Pozsgay2014,Brockmann2014a,Brockmann2014b,Brockmann2017}. It can be proved that the boundary state $\ket{{\cal{D}}_{\frac{1}{2}}}$ yields $g_{\cal{D}_{\frac{1}{2}}}=\sqrt[4]{3}$~\cite{Append}.

Now we turn to lattice models with periodic boundary conditions. For the ULS (AKLT) chain with $L$ sites, the ground state is denoted as $\ket{\psi_{0}(L)}$ ($\ket{\mathrm{AKLT}(L)}$). In principle, any state in the Haldane phase could serve as the lattice version of the boundary state (e.g., the ground state of $H(\gamma)$ for any $-1<\gamma<1$). Remarkably, the overlap can be computed most easily when the AKLT state is used, thanks to recent developments of intergrability techniques~\cite{Leeuw2016,Mestyan2017,Pozsgay2019}. The AKLT state is actually an ``integrable boundary state" for the ULS chain in the sense that the overlap $\langle\psi_0(L)|\mathrm{AKLT}(L)\rangle$ can be expressed in terms of the Bethe roots of the ULS chain. This allows for efficient computation of its numerical value. Furthermore, we analytically derive an {\em exact} asymptotics
\begin{align}
    \ln\left|\langle\psi_{0}(L)|\mathrm{AKLT}(L)\rangle\right| \approx -\alpha L + \ln g_{\cal{D}_{\frac{1}{2}}} + \cdots
\label{eq:MPS-ovlp-asym}
\end{align}
for the $L \rightarrow \infty$ limit using the method of nonlinear integral equations~\cite{Append}. Here $\alpha\approx 0.13$, $\ln g_{\cal{D}_{\frac{1}{2}}} = \frac{1}{4}\ln 3 \approx 0.275$ is the expected AL entropy, and the ellipsis denotes finite-size corrections that vanish in the thermodynamic limit. The length $L$ is chosen as multiples of $6$ in these calculations~\footnote{There are two reasons for this choice: 1) the determinant formula for the overlap is only known when $L=0$ mod 6; 2) The ULS model has a unique ground state when $L=0$ mod 3.}.

\begin{figure}[ht]
\centering
\includegraphics[width=0.40\textwidth]{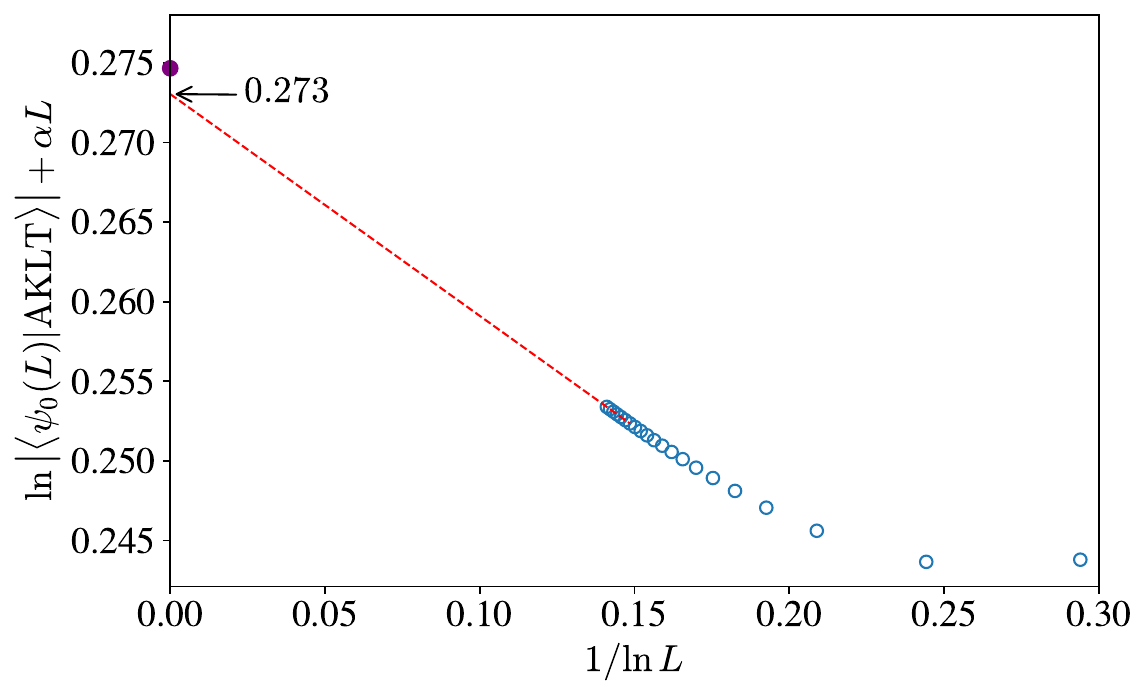}
\caption{Finite-size scaling of $\ln |\langle\psi_0(L)|\mathrm{AKLT}(L)\rangle | + \alpha L$ versus $1/\ln L$. The blue open circles represent numerical data obtained using the overlap formula. The results with $L\in [800,1200]$ are fitted by the red dashed line with equation $0.273 - 0.139/\ln L$. For reference, exact value of the AL entropy $\ln g_{\cal{D}_{\frac{1}{2}}} \approx 0.275$ is indicated by the purple dot.}
\label{fig:overlap}
\end{figure}

Although integrability enables direct extraction of the AL entropy in the large-$L$ limit, it is also of interest to examine the finite-size corrections in Eq.~\eqref{eq:MPS-ovlp-asym}, as they are expected to exhibit universal scaling behavior in the $L \gg 1$ regime. To this end, we consider the subtracted logarithmic overlap $\ln |\langle\psi_{0}(L)|\mathrm{AKLT}(L)\rangle | + \alpha L$. By numerically solving the Bethe roots, this quantity can be evaluated at quite large $L$ to high precision using the overlap formula~\cite{Append}. The results show a linear dependence on $1/\ln L$ at large $L$ as presented in Fig.~\ref{fig:overlap}. If data points in the range of $L\in [800,1200]$ are fitted, the extrapolated AL entropy has approximately $1\%$ relative error compared to the exact value. This logarithmic correction is a typical signature of marginally irrelevant perturbations due to lattice effects~\cite{Affleck1988,Cardy1986c}. It reinforces our previous analysis of the slow convergence observed in open chain simulations.

{\em Conclusions} --- In summary, we have demonstrated that an impurity coupled to a CFT can be screened by TDLs in this CFT even if its chiral primary fields fail to do so. This mechanism enriches our knowledge of BCFTs since it provides unconventional boundary conditions beyond the well-known Cardy states. A crucial step is to view different types of impurity as the edge modes of certain SPT states. For a SPT edge mode described by a specific projective class of the global symmetry, screening is expected to occur when there is a TDL labeled by the same projective class. In a spin-1 chain described by the SU(3)$_{1}$ CFT, one non-Cardy boundary state is constructed using conformal embedding, whose associated low-energy spectrum and AL entropy are verified by analytical and numerical calculations. The mechanism proposed here also applies to many other CFTs~\cite{Append}. While lattice spin models that realize some of them are known, the Hamiltonians may be quite complicated so numerical simulations are very challenging. We have studied a spin-2 model associated with the Spin(5)$_{1}$ CFT~\cite{Append}. Numerical results are consistent with theoretical predictions to some extent, but obvious discrepancies are observed. It will be interesting to study the stability of different boundary conditions and the renormalization group flows between them~\cite{Graham2004}. Another important future direction is to generalize this idea to higher dimensions.

\vspace{1em}

{\em Data availability} --- The data that support the findings of this article are openly available at~\cite{DataLink}.

\vspace{1em}

{\em Acknowledgments} --- M.C. acknowledges enlightening conversations with Sheng-Jie Huang, Chong Wang,  Dominic Else and Max Metlitski on related topics. H.-H.T. acknowledges helpful discussions with Philippe Lecheminant and Thomas Quella. This work was supported by the NNSF of China under grant No.~12174130 (Y.-H.W.), the Sino-German (CSC-DAAD) Postdoc Scholarship Program (Y.Z.), the Deutsche Forschungsgemeinschaft through project A06 of SFB 1143 under project No.~247310070 (H.-H.T.), and NSF under award number DMR-1846109 (M.C.).

\bibliography{ReferConde}

\begin{thebibliography}{88}
\bibitem [{App()}]{Append}%
  \BibitemOpen
  \href@noop {} {\bibinfo  {journal} {See the Appendices for more discussions
  on TDLs, non-Cardy boundary states from conformal embedding, details about
  numerical simulations of the ULS spin chains, detailes about the
  integrability techniques that are used to extract the Affleck-Ludwig entropy,
  and one more example that illustrates impurity screening by TDLs}\
  }\BibitemShut {NoStop}%
\end{thebibliography}

\clearpage
\onecolumngrid

\setcounter{figure}{0}
\setcounter{table}{0}
\setcounter{equation}{0}
\renewcommand{\thefigure}{A\arabic{figure}}
\renewcommand{\thetable}{A\arabic{table}}
\renewcommand{\theequation}{A\arabic{equation}}

\section{Appendix A: Brief review of topological defect lines}

For a given (1+1)$d$ CFT, its topological defect lines (TDLs) are line operators that commute with the stress tensor $T(z)$ on the spacetime worldsheet. In other words, such lines can be freely deformed. In the Hamiltonian formalism, a TDL corresponds to a generalized symmetry operator acting on the CFT Hilbert space. The ordinary global symmetry transformations represented by unitary operators are examples of invertible TDLs. Two TDLs $\cal{D}$ and $\cal{D}'$ can be fused to generate another TDL $\cal{D}\times\cal{D}'$. In the operator language, fusion simply means the multiplication of two operators. Another useful operation is the direct sums of TDLs. These two operations make the TDLs into a fusion ring that generalizes the group structure of ordinary symmetry. In particular, there is a set of {\em simple} TDLs whose members cannot be further decomposed into other TDLs. We denote these simple TDLs as $\cal{D}_{a},\cal{D}_{b},\cdots$. The fusion ring structure is dictated by the fusion rules
\begin{eqnarray}
    \cal{D}_{a} \times \cal{D}_{b} = \sum_{c} N^{c}_{ab} \cal{D}_{c}
\end{eqnarray}
with integer coefficients $N^{c}_{ab} \geq 0$. More details about the algebraic structure of TDLs can be found in Refs.~\cite{Chang:2018iay,Bhardwaj:2017xup}. A TDL can terminate at a defect operator. All defect operators for a given TDL $\cal{D}$ form a Hilbert space $\cal{H}_{\cal{D}}$. In view of the state-operator correspondence, $\cal{H}_{\cal{D}}$ is also the CFT Hilbert space on a circle which has $\cal{D}$ inserted along the time direction. While TDLs are usually discussed in the continuum, they can also be studied directly in lattice models. A large class of TDLs have been constructed in the ``anyon chain" models.

The general theory can be illustrated using the Ising CFT. It is well-known that this CFT has a self-duality symmetry that maps the order parameter $\sigma$ to the disorder parameter $\mu$. For a lattice realization (e.g. the transverse-field Ising model), this is implemented by the Kramers-Wannier duality transformation. We denote the TDL that corresponds to this self-duality as $\cal{N}$. The ordinary $\Z_{2}$ symmetry of the Ising CFT generates an invertible TDL called $\eta$. One can show that the fusion rules are 
\begin{eqnarray}
    && \cal{N}\times\cal{N}=1+\eta, \quad \eta\times\eta=1, \nonumber \\
    && \cal{N}\times\eta=\eta\times\cal{N} = \cal{N}.
\end{eqnarray}
An explaination of the rule $\eta\times\cal{N}=\cal{N}$ is helpful. Consider the action of $\cal{N}$ on the spin operator $\sigma(z)$ with $z$ being the complex coordinate. Pictorially, a TDL is dragged through the insertion of $\sigma(z)$. Under the duality, $\sigma(z)$ becomes the disorder operator $\mu(z)$, which must be attached to the end of a $\eta$ line. Since the action is local, the $\eta$ line has to end on the $\cal{N}$ line, which gives the fusion rule $\eta\times\cal{N}=\cal{N}$. This process is represented by the equation
\begin{equation*}
 \begin{tikzpicture}[baseline={([yshift=-.5ex]current bounding box.center)}]
    \draw[very thick,blue,rounded corners=4mm] (1,0)--(1,1)--(0.3,1.5)--(1,2) -- (1,3);
    \node at (1.2,1.5) [circle,fill,inner sep=1.5pt]{};
    \node at (1.4,1.2) [draw=none]{$\sigma(z)$};
    \node at (1.3,0.2) [draw=none]{${\cal N}$};
  \end{tikzpicture} \quad = \quad
\begin{tikzpicture}[baseline={([yshift=-.5ex]current bounding box.center)}]
    \draw[very thick,blue,rounded corners=4mm] (0.5,0)--(0.5,1)--(1.2,1.5)--(0.5,2) -- (0.5,3);
 \node at (-0.2,1.5) [circle,fill,inner sep=1.5pt]{};
    \node at (-0.25,1.2) [draw=none]{$\mu(z)$};
    \draw[thick, dashed] (-0.2,1.5)-- node[pos=0.4,above] {$\eta$}(1.06,1.5);
    \node at (0.8,0.2) [draw=none]{${\cal N}$};
  \end{tikzpicture}
\end{equation*}

For a rational CFT with diagonal partition function, there is a TDL associated with each chiral primary that are called Verlinde lines. In the Ising CFT, there are three chiral primary operators $1,\sigma,\psi$. We can identify $\cal{N}$ with the Verlinde line $\cal{D}_{\sigma}$ and $\eta$ with $\cal{D}_{\psi}$. It is also possible to construct non-Verlinde TDLs from conformal embedding as explained below.

\section{Appendix B: Trivialization of symmetry-protected topological phase stacking by TDLs}

Let us consider a CFT with global symmetry $G$ and TDL $\cal{D}$. In subsequent discussions, we use group elements to label TDLs of the $G$ symmetry. It is assumed that $\cal{D}$ commutes with $G$ in the sense that 
\begin{eqnarray}
    \cal{D} \times g = g \times \cal{D}, \quad \forall g \in G.
\end{eqnarray}
The defect Hilbert space $\cal{H}_{\cal{D}}$ transforms projectively under $G$. The projective class $\omega$ can be determined by the following relation:
\begin{equation}
\begin{tikzpicture}[baseline={([yshift=-.5ex]current bounding box.center)}]
    \def\height{3}    
    \def\radius{1}    
  
    \draw (0,\height) ellipse (\radius cm and 0.4*\radius cm);
    \draw[dashed] (\radius,0) arc(0:180:\radius cm and 0.4*\radius cm);
  \draw (-\radius,0) arc(180:360:\radius cm and 0.4*\radius cm);
 
    \draw (-\radius,\height) -- (-\radius,0);
    \draw (\radius,\height) -- (\radius,0);

    \draw[very thick] (0,-0.4*\radius) -- (0,\height-0.4*\radius);

    \draw[blue, thick, dashed] (\radius,1) arc(0:180:\radius cm and 0.4*\radius cm);
  \draw[thick, blue] (-\radius,1) arc(180:360:\radius cm and 0.4*\radius cm);
 
  \draw[blue, thick, dashed] (\radius,2) arc(0:180:\radius cm and 0.4*\radius cm);
  \draw[thick, blue] (-\radius,2) arc(180:360:\radius cm and 0.4*\radius cm);
 
  \node[below] at (0,-0.4*\radius) [draw=none]{$\cal{D}$};
\node[right] at (\radius,1) [draw=none]{$g$};
\node[right] at (\radius,2) [draw=none]{$h$};
\end{tikzpicture}
\:= \omega_{\cal{D}}(g,h)\:\:
\begin{tikzpicture}[baseline={([yshift=-.5ex]current bounding box.center)}]
    \def\height{3}    
    \def\radius{1}    
  
    \draw (0,\height) ellipse (\radius cm and 0.4*\radius cm);
    \draw[dashed] (\radius,0) arc(0:180:\radius cm and 0.4*\radius cm);
  \draw (-\radius,0) arc(180:360:\radius cm and 0.4*\radius cm);
 
    \draw (-\radius,\height) -- (-\radius,0);
    \draw (\radius,\height) -- (\radius,0);

    \draw[very thick] (0,-0.4*\radius) -- (0,\height-0.4*\radius);
 
  \draw[blue, thick, dashed] (\radius,1.5) arc(0:180:\radius cm and 0.4*\radius cm);
  \draw[thick, blue] (-\radius,1.5) arc(180:360:\radius cm and 0.4*\radius cm);
 \node[below] at (0,-0.4*\radius) [draw=none]{$\cal{D}$};
 \node[right] at (\radius,1.5) [draw=none]{$gh$};
\end{tikzpicture}
\end{equation}
Here the horizontal lines represent $G$ symmetry defect lines. This relation may be viewed as a kind of ``mixed anomaly" between the TDL $\cal{D}$ and the invertible symmetry $G$.

Now we perform a space-time rotation to obtain
\begin{equation}
\begin{tikzpicture}[baseline={([yshift=-.5ex]current bounding box.center)}]
    \def\height{2}    
    \def\radius{1}    
  
    \draw (0,\height) ellipse (\radius cm and 0.4*\radius cm);
    \draw[dashed] (\radius,0) arc(0:180:\radius cm and 0.4*\radius cm);
  \draw (-\radius,0) arc(180:360:\radius cm and 0.4*\radius cm);
 
    \draw (-\radius,\height) -- (-\radius,0);
    \draw (\radius,\height) -- (\radius,0);

    \draw[blue, thick] (0.35,-0.4*\radius+0.02) node[below]{$h$} -- (0.35,\height-0.4*\radius+0.02);
\draw[blue, thick] (-0.35,-0.4*\radius+0.02) node[below]{$g$} -- (-0.35,\height-0.4*\radius+0.02);

      \draw[very thick, dashed] (\radius,1) arc(0:180:\radius cm and 0.4*\radius cm);
  \draw[very thick] (-\radius,1) arc(180:360:\radius cm and 0.4*\radius cm);
 \node[right] at (\radius,1) [draw=none]{$\cal{D}$};
\end{tikzpicture}
\:= \omega_{\cal{D}}(g,h)\:\:
\begin{tikzpicture}[baseline={([yshift=-.5ex]current bounding box.center)}]
    \def\height{2}    
    \def\radius{1}    
  
    \draw (0,\height) ellipse (\radius cm and 0.4*\radius cm);
    \draw[dashed] (\radius,0) arc(0:180:\radius cm and 0.4*\radius cm);
  \draw (-\radius,0) arc(180:360:\radius cm and 0.4*\radius cm);
 
    \draw (-\radius,\height) -- (-\radius,0);
    \draw (\radius,\height) -- (\radius,0);

    \draw[blue, thick] (0,-0.4*\radius) node[below]{$gh$} -- (0,\height-0.4*\radius);

      \draw[very thick, dashed] (\radius,1) arc(0:180:\radius cm and 0.4*\radius cm);
  \draw[very thick] (-\radius,1) arc(180:360:\radius cm and 0.4*\radius cm);
 \node[right] at (\radius,1) [draw=none]{$\cal{D}$};
\end{tikzpicture}
\end{equation}
This equation can be used to define a $G$ symmetry-protected topological (SPT) phase: namely, fusing two $G$ defects may produce a nontrivial phase factor. If $\cal{D}$ is applied on a $G$-symmetric boundary state, we obtain a new boundary state whose projective class is modified by $\omega_{\cal{D}}$. On the other hand, the TDL is a generalized symmetry of the CFT. In particular, its action on the CFT ground state $\ket{0}$ yields
\begin{eqnarray}
    \cal{D}\ket{0} = \braket{\cal{D}}\ket{0}.
\end{eqnarray}
with $\braket{\cal{D}}>0$ being the expectation value of $\cal{D}$. In other words, the ground state is invariant under the action of $\cal{D}$. Applying $\cal{D}$ on any gapped $G$-invariant state has the same effect as stacking a $G$ SPT state labeled by $\omega_{\cal{D}}$, so we conclude that stacking a SPT with $\omega_{\cal{D}}$ on the CFT does not lead to a distinct symmetry-enriched CFT.

\section{Appendix C: Boundary states from conformal embedding}

All boundary states of a rational CFT that preserve (half of) the chiral algebra are of the Cardy type. However, non-Cardy boundary states with reduced symmetry may also exist. Here we explain how to construct them using conformal embedding. For a pair of chiral rational CFTs $\cal{R}$ and $\cal{R}'$, if the latter is obtained from the former by an extension of the chiral algebra, there is a conformal embedding $\cal{R}\subset\cal{R}'$. 

It is helpful to focus on the algebraic data of the CFT. A chiral CFT is associated with a modular tensor category (MTC), which is mathematically the representation category of the chiral algebra. We denote the MTCs associated with $\cal R$ and $\cal R'$ by $\cal{C}$ and $\cal{C}'$, respectively.  In this language, chiral algebra extension means $\cal{C}'$ is obtained from $\cal{C}$ by anyon condensation.  Mathematically, the condensate is described by a separable, commutative Frobenius object $A$ in $\cal{C}$. The interface between $\cal{C}$ and $\cal{C}'$ can be endowed with a fusion category $\cal{T}$, which contains both the deconfined anyons (i.e. $\cal{C}'$) and the confined defects. All simple objects in $\cal{T}$ can be lifted to $\cal{C}$. More precisely, the lifting of $\alpha\in\cal{T}$ is
\begin{eqnarray}
    l(\alpha) = \sum_{a\in\cal{C}} n_{{\alpha}a} \; a, \quad n_{{\alpha}a} \in \mathbb{Z}^{{\geq}0}.
\end{eqnarray}
In particular, the lifting of the vacuum in $\cal{T}$ is the algebra object $A$. A subcategory of $\cal{T}$ which contains all deconfined anyons becomes $\cal{C}'$, and the remaining ones become topological defects in $\cal{C}'$. If the characters of the CFT $\cal{R}$ are $\chi_{a}$, then the characters of the CFT $\cal{R}'$ can be expressed as
\begin{eqnarray}
    \chi_{\alpha} = \sum_{a\in\cal{C}} n_{{\alpha}a} \; \chi_{a}, \quad \alpha\in\cal{C}'.
\end{eqnarray}
Each object $\alpha$ in $\cal{T}$ defines a boundary state for $\cal{R}'$ and those belong to $\cal{C}'$ correspond to the Cardy states. For a cylinder with boundary states $\alpha,\beta$ at the two ends, its partition function is~\cite{HuangSJ2025} 
\begin{eqnarray}
    \cal{Z}_{\alpha,\beta} = \sum_{\gamma\in\cal{T}} N^\gamma_{\alpha\beta} \chi_{\gamma}.
\end{eqnarray}

Let us turn to specific examples. For the SU(3)$_{1}$ CFT, non-Cardy boundary states are constructed using the conformal embedding SU(3)$_{1}\subset$ SU(2)$_{4}$. From the viewpoint of topological quantum field theory, SU(3)$_{1}$ can be obtained from SU(2)$_{4}$ by condensing the algebra $0+2$. The primaries of SU$(3)_{1}$ and SU(2)$_{4}$ can be identified as
\begin{eqnarray}
    \mb{0} \simeq 0+2, \quad \mb{3} \simeq 1_{+}, \quad \bar{\mb{3}} \simeq 1_{-}, 
\end{eqnarray}
where $1_\pm$ means that the spin-1 field splits into $\mb{3}$ and $\bar{\mb{3}}$. In addition, there is a TDL $\cal{D}_{\frac{1}{2}} = \frac{1}{2} + \frac{3}{2}$ with fusion rules
\begin{eqnarray}
    \cal{D}_{\frac{1}{2}} \times \cal{D}_{\frac{1}{2}} = \mb{0}+\mb{3}+\bar{\mb{3}}, \quad \cal{D}_{\frac{1}{2}} \times \mb{3} = \cal{D}_{\frac{1}{2}} \times \bar{\mb{3}} = \cal{D}_{\frac{1}{2}}.
\end{eqnarray}

\section{Appendix D: Numerical simulations of the ULS spin chains}

As displayed in Fig.~1, there are two equivalent ways to represent our system. The spin-1/2 impurities in Fig.~1 (a) is connected to two sites of the ULS chain via Heisenberg terms. For the left edge, the coupling term is
\begin{eqnarray}
J_{1} \mb{S}_{\frac{1}{2}} \cdot \mb{S}_{1} + J_{2} \mb{S}_{\frac{1}{2}} \cdot \mb{S}_{2},
\label{eq:EdgeCouple}
\end{eqnarray}
where $\mb{S}_{1/2}$ is the spin-1/2 impurity and $\mb{S}_{1}$ ($\mb{S}_{2}$) is the first (second) site of the ULS chain. The coupling at the right edge (if there are two impurities) is similar. If $J_{1} \gtrsim 5$ and $J_{2}=0$, the results have negligible dependence on $J_{1}$. By tuning $J_{2}$ to a suitable negative value, finite-size effects can be suppressed to some extent. The data in the main text corresponds to $J_{1}=10,J_{2}=-5$.

\begin{figure}[ht]
\centering
\includegraphics[width=0.85\textwidth]{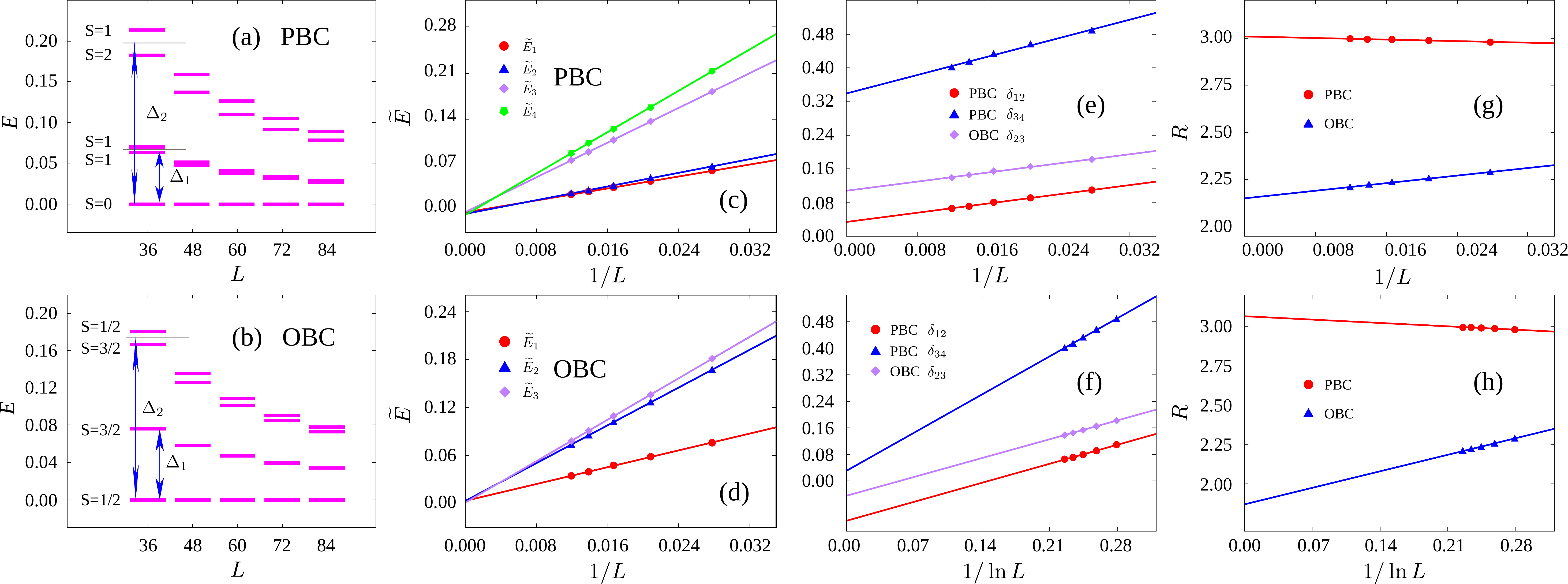}
\caption{Numerical results on the ULS-AKLT chains. (a,b) Energy spectra for the cases with PBC and OBC. The ground-state energy is shifted to zero. The total spin of each level is indicated. The avergaed spacings defined in the text are illustrated. (c,d) Finite-size scaling analysis of the energy spacings $\widetilde{E}_{i}=E_{i}-E_{0}$. (e,f) Finite-size scaling analysis of the energy splittings $\delta_{12},\delta_{34}$ for PBC and $\delta_{23}$ for OBC. (g,h) Finite-size scaling analyis of the ratios $R=\Delta_{2}/\Delta_{1}$.}
\label{fig:ULSAKLT}
\end{figure}

The model in Fig.~1 (b) is directly related to theoretical analysis that relies on SPT states. The system has a CFT part with $L$ spins and a SPT part with $L_{s}$ spins. It is favorable to set the SPT part at the AKLT point for its short correlation length. The full Hamiltonian is
\begin{eqnarray}
H = H_{\rm CFT} + H_{\rm SPT} + H_{\rm BC},
\end{eqnarray}
where $H_{\rm CFT} = H(\gamma=1)$ contains the indices $j\in \left[ 1,L-1 \right]$, $H_{\rm SPT} = 10 \, H(\gamma=\frac{1}{3})$ contains the indices $j\in \left[ L , L+L_{s}-1 \right]$, and $H_{\rm BC}$ specifies the boundary condition. The prefactor $10$ in $H_{\rm SPT}$ creates a large gap such that the CFT part only has non-negligible coupling with the AKLT edge modes. If the system has periodic boundary condition (PBC), the last and first sites are connected by 
\begin{eqnarray}
H_{\rm BC}=10 \left[ \mb{S}_{L+L_{s}} \cdot \mb{S}_{1} + \frac{1}{3} \left( \mb{S}_{L+L_{s}} \cdot \mb{S}_{1})^{2} \right) \right].
\end{eqnarray}
If the system has open boundary conditions (OBC), the open end of the AKLT part is connected to a spin-1/2 by 
\begin{eqnarray}
H_{\rm BC}=10 \;  \mb{S}_{L+L_{s}} \cdot \mb{S}_{\frac{1}{2}}, 
\end{eqnarray}
which removes an undesirable degeneracy caused by the SPT edge mode. Numerical results for this model are displayed in Fig.~\ref{fig:ULSAKLT}. We analyze the data using the same procedure described in the main text. The energy spacings $\widetilde{E}_{i}=E_{i}-E_{0}$ decay linearly with $L$ as shown in Fig.~\ref{fig:ULSAKLT} (c,d). To confirm that certain levels would collapse in the thermodynamic limit, the splittings $\delta_{12}=(E_{2}-E_{1})/\widetilde{E}_{1},\delta_{34}=(E_{4}-E_{3})/\widetilde{E}_{1}$ for PBC and $\delta_{23}=(E_{3}-E_{2})/\widetilde{E}_{1}$ for OBC are plotted in Fig.~\ref{fig:ULSAKLT} (e,f). As in the main text, better fitting is achieved when the horizontal axis is $1/\ln{L}$. We define averaged spacings $\Delta_{1}=(E_{1}+E_{2})/2-E_{0},\Delta_{2}=(E_{3}+E_{4})/2-E_{0}$ for PBC and $\Delta_{1}=E_{1}-E_{0},\Delta_{2}=(E_{2}+E_{3})/2-E_{0}$ for OBC. The ratios $R=\Delta_{2}/\Delta_{1}$ are presented in Fig.~\ref{fig:ULSAKLT} (g,h) and also given in Table~\ref{TableS1}. It is evident that finite-size effects are stronger here. While the fitting versus $1/\ln{L}$ has a better quality for OBC, the extrapolated value deviates substantially from $2$.

\begin{table}[ht]
\begin{tabular}{c|ccccc}
\hline
\hline
$L$         & 36 & 48 & 60 & 72 & 84 \\
            \cline{2-4}          
\hline
PBC             &  2.980  &  2.988  &  2.992  &  2.994  &  2.996 \\
OBC             &  2.289  &  2.257  &  2.236  &  2.221  &  2.209 \\
\hline
\hline
\end{tabular}
\caption{Numerical values of $R=\Delta_{2}/\Delta_{1}$ in the ULS-AKLT chains.}
\label{TableS1}
\end{table}

\section{Appendix E: Extracting the Affleck-Ludwig entropy}

The spherical basis of the $\mathrm{SO}(3)$ fundamental representation, denoted by $|a\rangle$ $(a=0,1,2)$, is related to the standard $S^z$-basis of spin-1 as follows:
\begin{align}
    |0\rangle = \frac{-1}{\sqrt{2}} \left(|S^z =1\rangle - |S^z = -1\rangle\right)\, , \quad
    |1\rangle = \frac{i}{\sqrt{2}} \left(|S^z =1\rangle + |S^z = -1\rangle\right)\, , \quad
    |2\rangle = |S^z = 0\rangle\, .
\end{align}
In the spherecial basis, the spin-1 AKLT state in a periodic chain with $L$ sites is written as
\begin{align}
    |\mathrm{AKLT}(L)\rangle = \frac{1}{\sqrt{\mathcal{N}}} \sum_{a_1, \ldots, a_L = 0,1,2} \mathrm{Tr} (\sigma^{a_1} \cdots \sigma^{a_L} ) |a_1\ldots a_L\rangle \, ,
\label{eq:AKLT-SM}
\end{align}
where $\sigma^0 \equiv \sigma^x$, $\sigma^1 \equiv \sigma^y$, $\sigma^2 \equiv \sigma^z$ are Pauli matrices and $\mathcal{N} = 3^L + 3 (-1)^L $ is the normalization factor. 

By identifying the spherical basis with the $\mathrm{SU}(3)$ spin basis, the spin-1 AKLT state was found to be an integrable matrix product state (MPS) of the $\mathrm{SU}(3)$ ULS spin chain~\cite{Leeuw2016,Pozsgay2019}. In the following discussion, we assume that the chain length satisfies $\mathrm{mod}(L,6)= 0$, so that the ground state of the ULS chain, denoted by $|\psi_0(L)\rangle$, is unique and characterized by a set of Bethe roots with a paired structure: $\{u_k^{(a)}\} = \{u_{k_+}^{(a)}\}\cup \{-u_{k_+}^{(a)}\}$, with $a=1,2$ and $k=1,\ldots,M_a$, where $M_1 = 2L/3$ and $M_2=L/3$. These Bethe roots satisfy the nested Bethe Ansatz equations (BAEs):
\begin{align}
    \phi^{(a)}_L (u^{(a)}_k) = \frac{2\pi}{L}\left(k-\frac{M_a-1}{2}\right) \, ,
    \label{eq:BAEs-SM}
\end{align}
where the counting functions are defined as
\begin{align}
    \phi_L^{(a)} (u) &= \vartheta_1(u)\delta_{a,1} 
    + \frac{1}{L}\sum_{l=1}^{M_{a-1}}\vartheta_1(u-u_l^{(a-1)}) - \frac{1}{L}\sum_{l=1}^{M_a}\vartheta_2(u-u_l^{(a)}) 
    + \frac{1}{L}\sum_{l=1 }^{M_{a+1}}\vartheta_1(u-u_l^{(a+1)})
\label{eq:count-func-SM}
\end{align}
with $\vartheta_p(u)=2\arctan(2u/p)$ for $p=1,2$. In Eq.~\eqref{eq:count-func-SM}, we take $a=1,2$ and define $M_0 = M_3 = 0$ as a convention. The overlap between the AKLT state and the ULS ground state is given by~\cite{Leeuw2016}
\begin{align}
    \left|\langle \psi_{0}(L) | \mathrm{AKLT}(L) \rangle\right| =\frac{2}{\sqrt{\mathcal{N}}}\sqrt{\prod_{a=1,2}\prod_{k_+ =1}^{M_a/2} \frac{(u^{(a)}_{k_+})^2 + 1/4}{(u^{(a)}_{k_+})^2}}\sqrt{\frac{\mathrm{det} G_L^+}{\mathrm{det} G_L^-}} \, ,
\label{eq:AKLT-ovlp-SM}
\end{align}
where $G_L^\pm$ are factorized Gaudin matrices whose indices run over the positive Bethe roots:
\begin{align}
    [G_L^\pm]_{ab;k_+l_+} = 2\pi L\rho^{(a)}_L (u_{k_+}^{(a)}) \delta_{ab}\delta_{k_+l_+} 
    + K_{ab}(u_{k_+}^{(a)}-u_{l_+}^{(b)}) \pm K_{ab}(u_{k_+}^{(a)}+u_{l_+}^{(b)})\,,
\label{eq:Gaudin-mat-SM}
\end{align}
with auxiliary functions defined as
\begin{align}
    \rho_L^{(a)} (u) &= \frac{1}{2\pi}\frac{\mathrm{d}}{\mathrm{d}u}\phi^{(a)}_L (u) \,, \nonumber\\
    K_{ab}(u) &= \delta_{ab}\frac{\mathrm{d}}{\mathrm{d}u}\vartheta_2 (u) - (\delta_{a+1,b}+\delta_{a-1,b})\frac{\mathrm{d}}{\mathrm{d}u}\vartheta_1 (u)\,.
\label{eq:aux-func-SM}
\end{align}

To analyze the asymptotic behavior in the large-$L$ limit, we decompose the logarithm of the overlap in Eq.~\eqref{eq:AKLT-ovlp-SM} into three terms:
\begin{align}
    \ln\left| \langle\psi_{0}(L) | \mathrm{AKLT}(L) \rangle\right| = \left(\ln 2 -\frac{1}{2}\ln\mathcal{N} \right) + \frac{1}{2}\ln \mathcal{P}(L) + \frac{1}{2}\ln \left(\frac{\mathrm{det}G_N^+}{\mathrm{det}G_N^-}\right)
\label{eq:logovlp-SM}
\end{align}  
with
\begin{align}
    \mathcal{P}(L) \equiv  \prod_{a=1,2}\prod_{k_+ =1}^{M_a/2} \frac{(u^{(a)}_{k_+})^2 + 1/4}{(u^{(a)}_{k_+})^2} \, .
\end{align}
The first term in Eq.~\eqref{eq:logovlp-SM} admits the asymptotic expansion 
\begin{align}
    \ln 2 -\frac{1}{2}\ln\mathcal{N} = - \frac{\ln 3 }{2} L + \ln 2 + \mathcal{O} (e^{-L\ln 3}) \, ,
    \label{eq:1st-term-asym-SM}
\end{align}
while the second and third terms explicitly depend on the Bethe roots.

\begin{figure}[ht]
\centering
\includegraphics[width=0.55\textwidth]{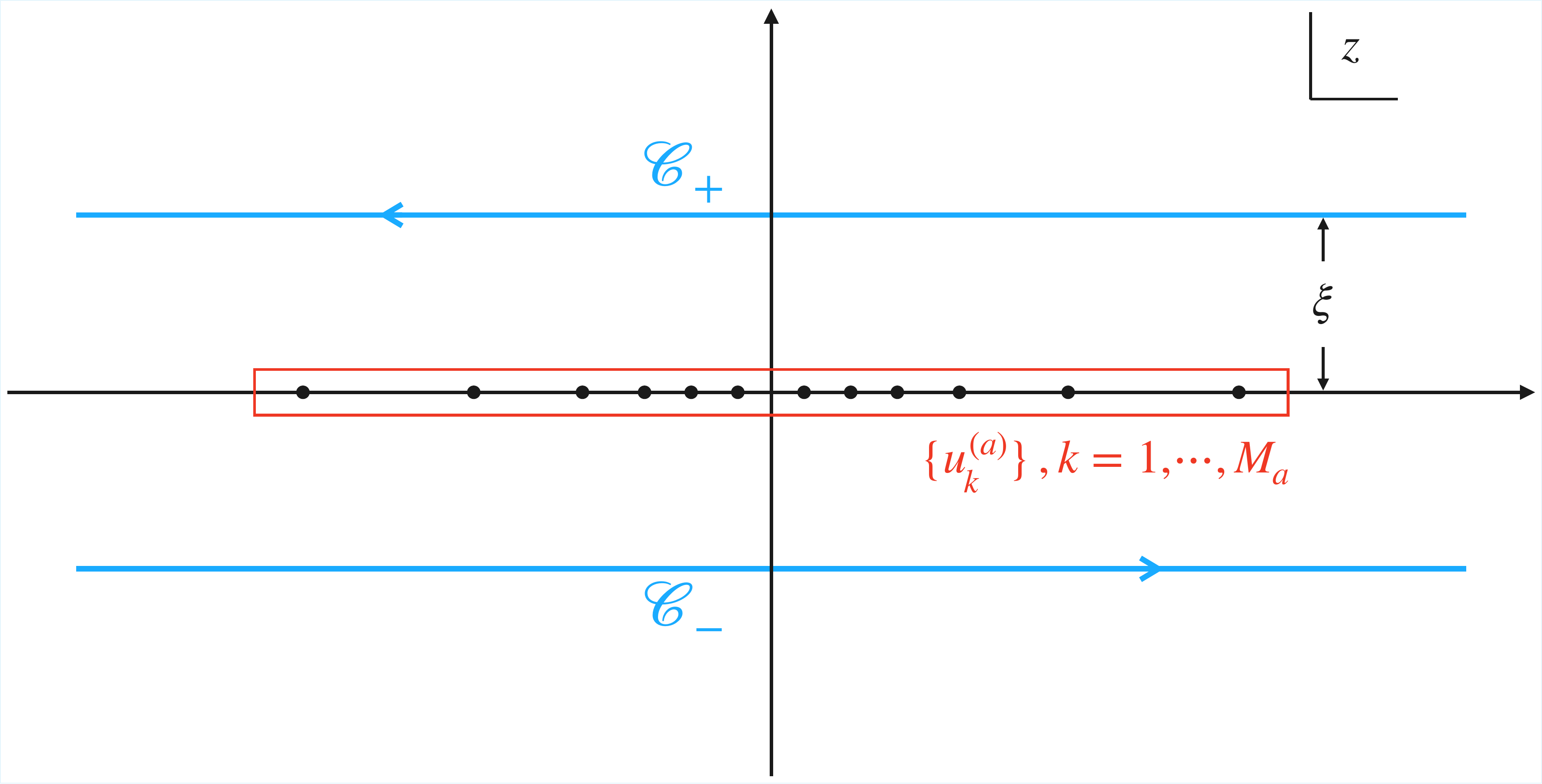}
\caption{The integral contour $\mathscr{C} = \mathscr{C}_+ + \mathscr{C}_-$ in Eq.~\eqref{eq:contour-SM}, with $\xi \in (0,\frac{1}{2})$.}
\label{fig:contour-SM}
\end{figure}

To obtain the asymptotic expansion for the second and third terms, we apply the technique of non-linear integral equations (NLIEs)~\cite{Klumper1991,Klumper1993,Destri1992,Destri1995,Brockmann2017,Pozsgay2018}. The main idea of the NLIEs is to convert sums over Bethe roots of the form $\sum_k h(u^{(a)}_k)$, where $h(u)$ is an analytic function, into contour integrals in the complex plane. Notably, the counting functions $\phi^{(a)}_N(z)$ [Eq.~\eqref{eq:count-func-SM}] are analytic in the strip $\{z|\mathrm{Im}\, z \in (-\frac{1}{2},\frac{1}{2})\}$ and fully encode the Bethe roots via the BAEs. This allows such sums to be expressed as
\begin{align}
    \sum_{k=1}^{M_a} h(u^{(a)}_k) = \frac{1}{2\pi i}\oint_{\mathscr{C}} h(z) \mathrm{d}\ln \left(1 + e^{iN\phi^{(a)}_L(z)}\right)\,,\quad a=1,2\, ,
    \label{eq:contour-SM}
\end{align}
where the integral contour $\mathscr{C}$ is chosen to enclose all Bethe roots $\{u_k^{(a)}\}$ on the real axis. For convenience, we take the contour $\mathscr{C}$ to encircle the entire real line, as illustrated in Fig.~\ref{fig:contour-SM} with $0 < \xi < \frac{1}{2}$. 

Now, expressing the sums in Eq.~\eqref{eq:count-func-SM} as contour integrals, we obtain a set of coupled NLIEs of counting functions:
\begin{align}
    \phi_L^{(a)} (u) &= \vartheta_1(u)\delta_{a,1} \nonumber\\
    &+ \frac{1}{2\pi L i}\oint_{\mathscr{C}}\left[\vartheta_1(u-z) \mathrm{d}\ln \left(1 + e^{iL\phi^{(a-1)}_L(z)}\right) - \vartheta_2(u-z)\mathrm{d}\ln \left(1 + e^{iL\phi^{(a)}_L(z)}\right) + \vartheta_1(u-z)\mathrm{d}\ln \left(1 + e^{iL\phi^{(a+1)}_L(z)}\right)\right]  \, .
    \label{eq:count-func-contour-SM}
\end{align}
Integrating by parts and convolving with the kernel
\begin{align}
    G_{ab}(u) 
    &\equiv \left[\delta_{ab}\delta + \frac{1}{2\pi}K_{ab}\right]^{-1}(u) 
    = \frac{1}{2\pi} \int_{-\infty}^{+\infty} \mathrm{d}q \, e^{iq u} \tilde{G}_{ab} (q) \, ,\nonumber \\
    \tilde{G}_{ab} (q) 
    &= \frac{e^{|q|/2}}{3} \sum_{k=1,2}\frac{\sin (\pi k a/3)\sin(\pi k b/3)}{\cosh(q/2)-\cos(\pi k/3)} \, , \quad a,b = 1,2\, ,
\label{eq:Green-kernel-SM}
\end{align}
where $K_{ab}(u)$ has been defined in Eq.~\eqref{eq:aux-func-SM}, we arrive at
\begin{align}
    \phi_L^{(a)} (u) 
    = \phi^{(a)} (u)  + \frac{2}{L}\, \mathrm{Im} \int_{-\infty}^{+\infty} \mathrm{d}v \, [\delta_{ab}\delta -G_{ab}](u-v-i\xi)\ln \left(1+e^{iL\phi^{(b)}_L(v+ i\xi)}\right)  \, , \quad u\in\mathbb{R} \, ,\quad 0<\xi <\frac{1}{2} \, ,
\label{eq:count-func-NLIEs-SM}
\end{align}
with
\begin{align}
    \phi^{(a)} (u) = 2\arctan\left[\coth\left(\frac{\pi a}{6}\right)\tanh\left(\frac{\pi u}{3}\right)\right] \, ,\quad a=1,2\, .
\label{eq:count-func-thermo-SM}
\end{align}
As the soultion of the NLIEs [Eq.~\eqref{eq:count-func-NLIEs-SM}], it is not difficult to see that, in the thermodynamic limit, the counting functions are given by
\begin{align}
    \lim_{L\to\infty}\phi_L^{(a)} (u) = \phi^{(a)} (u)\, .
\end{align}
In fact, we can further examine the asymptotic behavior of the non-linear term appearing in the NLIEs [Eq.~\eqref{eq:count-func-NLIEs-SM}]. We are free to deform the contour by taking
\begin{align}
    \xi \equiv \epsilon \to 0^+\,,
\end{align}
so that the non-linear term reads
\begin{align}
    \mathrm{Im} \int_{-\infty}^{+\infty} \mathrm{d}v [\delta_{ab}\delta -G_{ab}](u-v-i\epsilon)\ln \left(1+e^{iL\phi^{(b)}_L(v+ i\epsilon)}\right)\, .
\end{align}
The thermodynamic limit corresponds to the double limit, $\lim_{\epsilon\to 0^+}\lim_{L\to\infty}$ (the order of two limits is important). Since $\phi^{(a)} (u+i\epsilon) = \phi^{(a)} (u) + 2\pi i\epsilon \rho^{(a)} (u) + \mathcal{O}(\epsilon^2)$, with $\rho^{(a)} (u) > 0$, the non-linear term vanishes in this double limit. This result can be generalized to any test function $h(u)$ that vanishes at infinity, i.e., $h(\pm \infty) \to 0$. For such functions, we have
\begin{align}
    \lim_{\epsilon\to 0^+}\lim_{L\to\infty} \int_{-\infty}^{+\infty} h(u) \ln\left[1 + e^{iL\phi^{(a)}_L(u+i\epsilon)}\right] \mathrm{d}u  = 0 \, .
    \label{eq:non-linear-thermo-SM}
\end{align}

We can use the same strategy to express the second and third terms in Eq.~\eqref{eq:logovlp-SM} as contour integrals. 

\begin{figure}[ht]
\centering
\includegraphics[width=0.55\textwidth]{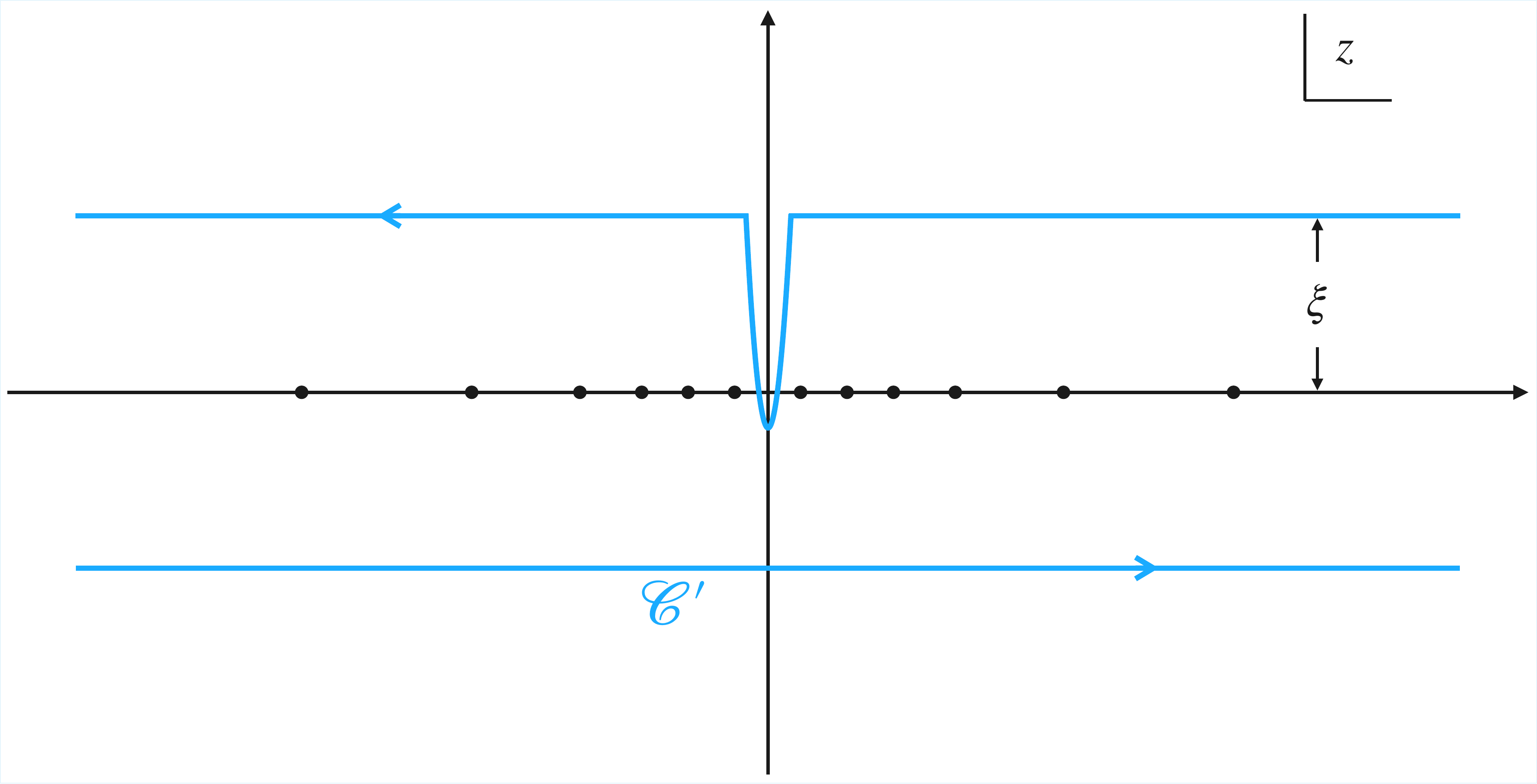}
\caption{The deformed integral contour $\mathcal{C}'$ in Eq.~\eqref{eq:log-pref-SM}, with $\xi \in (0,\frac{1}{2})$.}
\label{fig:deformed-contour-SM}
\end{figure}

For the second term in Eq.~\eqref{eq:logovlp-SM}, we write it as $\ln \mathcal{P}(L) = \sum_{a=1,2} \ln \mathcal{P}_a (L)$, with
\begin{align}
    \ln \mathcal{P}_a(L) 
    = \sum_{k=1}^{M_a} f(u^{(a)}_k) 
    = \frac{1}{2\pi i}\oint_{\mathscr{C}'} f(z) \mathrm{d}\ln\left[1+ e^{i L\phi^{(a)}_L (z)}\right]
    \label{eq:log-pref-SM}
\end{align}
and $f(u) = \frac{1}{2}\ln (\frac{u^2 + 1/4}{u^2})$. The contour $\mathscr{C}'$ is deformed to avoid the branch point of $f(u)$, as illustrated in Fig.~\ref{fig:deformed-contour-SM}. 

Integrating by parts and applying the residue theorem, we can express $\ln \mathcal{P}_a(L)$ as an integral over contour $\mathscr{C}$, along with the contribution from the residue:
\begin{align}
    \ln\mathcal{P}_a(N) 
    = -\frac{1}{2\pi i}\oint_\mathscr{C} f'(z) \ln\left[1+ e^{i L\phi^{(a)}_L (z)}\right] \mathrm{d}z - \ln 2 \, .
\label{eq:log-pref-2-SM}
\end{align}
Using the NLIEs of the counting functions [Eq.~\eqref{eq:count-func-NLIEs-SM}], the contour integral in Eq.~\eqref{eq:log-pref-2-SM} can be written as
\begin{align}
    &\quad -\frac{1}{2\pi i}\oint_\mathscr{C} f'(z) \ln\left[1+ e^{i L\phi^{(a)}_L (z)}\right] \mathrm{d}z \nonumber\\
    &= L \int_{-\infty}^{+\infty} f(u) \rho^{(a)} (u) \, \mathrm{d}u   + 2\,\mathrm{Re}\int_{0}^{+\infty} \mathrm{d}u \,[\delta_{ab}\delta -G_{ab}](-u-i\xi) \ln\left[1 + e^{iL\phi^{(b)}_L(u+i\xi)}\right]\nonumber\\
    &\quad +\frac{1}{\pi}\, \mathrm{Im} \int_{-\infty}^{+\infty} \mathrm{d}u \, f'(u+i\xi)\int_{-\infty}^{+\infty} \mathrm{d}v \, G_{ab}(u-v) \ln\left[1 + e^{iL\phi^{(b)}_L(v+i\xi)}\right]
     \, ,\quad 0<\xi <\frac{1}{2}\, ,
     \label{eq:log-pref-3-SM}
\end{align}
where 
\begin{align}
    \rho^{(a)} (u) = \lim_{L\to\infty} \rho_L^{(a)} (u)  = \frac{1}{2\pi} \frac{\mathrm{d}}{\mathrm{d} u} \phi^{(a)} (u) = \frac{1}{3}\left[\frac{\sin (\pi a/3)}{\cosh(2\pi u/3)-\cos (\pi a/3)}\right] 
    \label{eq:rho-thermo-SM}
\end{align}
is the auxiliary function $\rho_L^{(a)} (u)$ in the thermodynamic limit, interpreted as the Bethe root density. The first term in Eq.~\eqref{eq:log-pref-3-SM} contributes to the non-universal surface free energy, while the remaining non-linear terms vanish in the large-$L$ limit, as discussed in Eq.~\eqref{eq:non-linear-thermo-SM}. Therefore, the asymptotic expansion of $\ln \mathcal{P}(L)$ reads
\begin{align}
    \ln \mathcal{P}(L) = L \left[\sum_{a=1,2} \int_{-\infty}^{+\infty} f(u) \rho^{(a)} (u) \, \mathrm{d}u \right] -2\ln 2 + \cdots\, ,
    \label{eq:2nd-term-asym-SM}
\end{align}
where the ellipsis ``$\cdots$'' denotes higher-order corrections that vanish as $L\to \infty$.

To deal with the third term in Eq.~\eqref{eq:logovlp-SM}, we follow the approach outlined in Ref.~\cite{Pozsgay2018}. We begin by rewriting the determinant ratio as
\begin{align}
    \frac{\mathrm{det}G_L^+}{\mathrm{det}G_L^-} = \frac{\mathrm{det}(1-H_L^+)}{\mathrm{det}(1-H_L^-)}
\end{align}
with $H_L^\pm$ being matrices indexed by the full set of Bethe roots, defined as
\begin{align}
    H_L^\pm = \frac{P_L \pm Q_L}{2}
\end{align}
with
\begin{align}
    [P_L]_{ab;kl} &\equiv P_{ab} (u_k^{(a)},u_l^{(b)})= -i\frac{\mathfrak{a}^{(a)}_L(u_k^{(a)})}{{\mathfrak{a}^{(a)}_L}'(u_k^{(a)})} K_{ab}(u_k^{(a)}-u_l^{(b)}) \, ,\nonumber\\
    [Q_L]_{ab;kl} &\equiv Q_{ab} (u_k^{(a)},u_l^{(b)})= -i\frac{\mathfrak{a}^{(a)}_L(u_k^{(a)})}{{\mathfrak{a}^{(a)}_L}'(u_k^{(a)})} K_{ab}(u_k^{(a)}+u_l^{(b)})\, ,
\end{align}
where $\mathfrak{a}^{(a)}_L(u) = e^{iL\phi^{(a)}_L(u)}$ for $a=1,2$, and ${\mathfrak{a}_L^{(a)}}'(u)$ denotes the derivative with respect to $u$. Using the relations $Q_L P_L = P_L Q_L$ and $Q_L Q_L = P_L P_L$, the third term in Eq.~\eqref{eq:logovlp-SM} can be expanded as
\begin{align}
    \ln\left[\frac{\mathrm{det}G_L^+}{\mathrm{det}G_L^-}\right]  = -\sum_{q=1}^\infty \frac{1}{q}\mathrm{Tr}\left[(H^+_L)^q-(H^-_L)^q\right] = -\sum_{q=1}^\infty \frac{1}{q}\mathrm{Tr}[P_L^{q-1}Q_L]\, ,
\end{align}
which admits the contour integral representation:
\begin{align}
    \mathrm{Tr}[P_L^{q-1}Q_L] &=\sum_{a_1,\ldots,a_q}\left[\prod_{j=1}^{q} \frac{1}{2\pi i}\oint_{\mathscr{C}} \mathrm{d}\ln \left(1+\mathfrak{a}^{(a_j)}_L(z_j)\right) \right]P_{a_1a_2} (z_1,z_{2})\cdots P_{a_{q-1}a_q} (z_{q-1},z_{q}) Q_{a_qa_1} (z_q,z_1)\nonumber\\
    &= \sum_{a_1,\ldots,a_q}\left[\prod_{j=1}^{q} \oint_{\mathscr{C}} \frac{\mathrm{d}z_j}{2\pi }\frac{-\mathfrak{a}^{(a_j)}_L(z_j)}{1+\mathfrak{a}^{(a_j)}_L(z_j)} \right] K_{a_1,a_2}(z_1-z_2)\cdots K_{a_{q-1},a_q}(z_{q-1}-z_q)K_{a_q,a_1}(z_n+z_1) \, ,
\end{align}
where the integral contour $\mathscr{C}$ is depicted in Fig.~\ref{fig:contour-SM}. Deforming the contour by taking $\xi \equiv \epsilon \to 0^+$, only the integration along the lower contour $\mathscr{C}_-$ contributes in the thermodynamic limit, since
\begin{align}
    \lim_{\epsilon\to 0^+}\lim_{L\to\infty}\frac{\mathfrak{a}_L^{(a)}(u+i\epsilon)}{1+ \mathfrak{a}_L^{(a)}(u+i\epsilon)} =0 \, , \quad \lim_{\epsilon\to 0^+}\lim_{L\to\infty}\frac{\mathfrak{a}_L^{(a)}(u-i\epsilon)}{1+ \mathfrak{a}_L^{(a)}(u-i\epsilon)} =1 \, ,\quad a = 1,2\, ,
    \quad \forall u\in\mathbb{R}\, .
\end{align}
Therefore, we obtain
\begin{align}
    \lim_{L\to\infty} \mathrm{Tr}[P_L^{q-1}Q_L] 
    &= (-1)^q\sum_{a_1,\ldots,a_q} \left[\prod_{j=1}^q \int_{-\infty}^{+\infty}\frac{\mathrm{d}u_j}{2\pi }\right]
    K_{a_1,a_2}(u_1-u_2)\cdots K_{a_{q-1},a_q}(u_{q-1}-u_q)K_{a_q,a_1}(u_q+u_1) \nonumber \\
    &= \frac{(-1)^q}{2} \mathrm{Tr}\left[\left(\int_{-\infty}^{+\infty}\frac{\mathrm{d}u}{2\pi }K(u)\right)^q\right] \, ,
\end{align}
where the kernel integral evaluates to 
\begin{align}
    \int_{-\infty}^{+\infty}\frac{\mathrm{d}u}{2\pi} K_{ab}(u) 
    = \begin{pmatrix}
        1 & -1 \\ -1 & 1
    \end{pmatrix}_{ab} \, ,
\end{align}
whose eigenvalues are $\lambda_1 = 0$ and $\lambda_2 = 2$. This leads to
\begin{align}
    \lim_{L\to \infty} \ln\left[\frac{\mathrm{det}G_L^+}{\mathrm{det}G_L^-}\right]  = \frac{1}{2} \sum_{a=1,2} \ln (1+\lambda_a) = \frac{1}{2} \ln 3 \, .
    \label{eq:3rd-term-asym-SM}
\end{align}

Combining the results from Eqs.~\eqref{eq:1st-term-asym-SM}, \eqref{eq:2nd-term-asym-SM}, and \eqref{eq:3rd-term-asym-SM}, we obtain the asymptotic expansion of the logarithmic overlap between the AKLT state and the ULS ground state:
\begin{align}
    \ln\left| \langle\psi_{0}(L) | \mathrm{AKLT}(L) \rangle\right| = -\alpha L + \frac{1}{4}\ln 3 + \cdots\, ,
\end{align}
which gives the expected Affleck-Ludwig entropy
\begin{align}
    \ln g_{\mathcal{D}} = \frac{1}{4} \ln 3 \, ,
\end{align}
and the non-universal surface free energy density
\begin{align}
    \alpha = \frac{\ln 3}{2} + \sum_{a=1,2} \frac{\sin (\pi a/3)}{6}\int_0^\infty \mathrm{d}u\, \frac{\ln [u^2/(u^2+1/4)] }{\cosh(2\pi u/3)-\cos (\pi a/3)} \approx 0.13 \, .
\end{align}

\section{Appendix F: Spin-2 chain: Spin(5)$_1$ CFT}

This section discusses another example of impurity screening by TDLs in the Spin(5)$_1$ CFT. First we briefly review its properties. There are three primary fields: the identity $\mathbf{0}$, the SO(5) spinor representation $\sigma$, and the SO(5) vector representation $\psi$. In terms of the SO(3) subgroup, the spinor representation $\sigma$ carries spin-$3/2$, so the associated Cardy state can in principle screen half-integer spins. On the other hand, non-Cardy boundary states can be constructed using the conformal embedding ${\rm SU}(2)_{10}\subset {\rm Spin}(5)_{1}$. To obtain Spin(5)$_{1}$ from SU(2)$_{10}$, the chiral algebra of the latter should be extended using the SU(2) spin-3 chiral primary operator. This leads to the following identification of the primaries:
\begin{eqnarray}
    \mathbf{0} \simeq 0+3, \quad \sigma \simeq \frac{3}{2}+\frac{7}{2}, \quad \psi \simeq 2+5,
\end{eqnarray}
as well as three
TDLs $\cal{D}_{1,2,3}$
\begin{eqnarray}
    \cal{D}_{1} \simeq \frac{1}{2} + \frac{5}{2} + \frac{7}{2}, \quad \cal{D}_{2} \simeq \frac{3}{2} + \frac{5}{2} + \frac{9}{2}, \quad \cal{D}_{3} = 1+2+3+8.
\end{eqnarray}
They do not correspond to any ordinary global symmetry of the CFT but are non-invertible generalized symmetries~\cite{Chang:2018iay,McGreevy:2022oyu,Cordova:2022ruw}.
 
Among all the TDLs, ${\cal D}_1, {\cal D}_2$ and $\sigma$ have half-integer spins. These defects have fusion rules
\begin{eqnarray}
    \cal{D}_{1} \times \cal{D}_{1} = 1+\cal{D}_{3}, \quad \cal{D}_{1} \times \psi = \cal{D}_{2}, \quad \cal{D}_{1} \times \sigma = \cal{D}_{3}.
    \label{Spin5TDL}
\end{eqnarray}
Three boundary states can be constructed by fusing them with the Cardy state $\ket{\mathbf{0}}$.
The fact that $N^{1}_{\cal{D}_{1}, \cal{D}_{1}}=1$ shows that $\cal{D}_{1}$ is a simple line, and the same is true for $\cal{D}_{2}$.
For various choices of boundary conditions, the partition functions are
\begin{eqnarray}
    \cal{Z}_{\cal{D}_{1},\cal{D}_{1}} &=& \left( \chi_{0}+\chi_{3} \right) + \left( \chi_{1}+\chi_{2}+\chi_{3}+\chi_{4} \right) \nonumber \\
    &=& q^{-\frac{5}{48}} \left( 1+3q^{\frac{1}{6}}+5q^{\frac{1}{2}}+17q+\cdots \right), \nonumber \\
    \cal{Z}_{\cal{D}_{2},\cal{D}_{2}} &=& \cal{Z}_{\cal{D}_{1},\cal{D}_{1}}, \nonumber \\
    \cal{Z}_{\cal{D}_{1},\mathbf{0}} &=& \chi_{\frac{1}{2}}+\chi_{\frac{5}{2}}+\chi_{\frac{7}{2}} \nonumber \\
    &=& q^{-\frac{1}{24}} \left( 2+6q^{\frac{2}{3}}+6q+8q^{\frac{5}{4}}+\cdots \right), \nonumber \\
    \cal{Z}_{\cal{D}_{2},\mathbf{0}} &=& \chi_{\frac{3}{2}}+\chi_{\frac{5}{2}}+\chi_{\frac{9}{2}} \nonumber \\
    &=& q^{\frac{5}{24}} \left( 4+6q^{\frac{5}{12}}+12q+18q^{\frac{17}{12}}+\cdots \right), \nonumber \\
    \cal{Z}_{\sigma,\sigma} &=& q^{-\frac{5}{48}} \left( 1+5q^{\frac{1}{2}}+10q+15q^{\frac{3}{2}}+\cdots \right), \nonumber \\
    \cal{Z}_{\sigma,\mathbf{0}} &=& q^{\frac{5}{24}} \left( 4+20q+60q^{2}+\cdots \right).
\end{eqnarray}
The Cardy state $\ket{\sigma}$ preserves the full SO(5) symmetry, but the non-Cardy states $\ket{\cal{D}_{1,2}}$ only preserve the SO(3) subgroup. All of them can in principle screen a spin-1/2 impurity according to the general theory. Predictions for low-energy spectra of an open chain are as follows. If both ends have $\ket{\cal{D}_{1}}$, from the partition function ${\cal Z}_{{\cal D}_1, {\cal D}_1}$ we see that the first three levels should have degeneracy $1,3,5$ and the energy spacing ratio $R=\Delta_{2}/\Delta_{1}=3$. If one end has $\ket{\cal{D}_{1}}$ and the other has $\ket{\mathbf{0}}$, according to the partition function ${\cal D}_{{\cal D}_1, \mathbf{0}}$, the first three levels have degeneracy $2,6,6$ and the energy spacing ratio $R=\Delta_{2}/\Delta_{1}=3/2$. If the boundary state is $\ket{\sigma}$ or $\ket{\cal{D}_{2}}$, the patterns of degeneracy and gap ratios would be very different.

The Spin(5)$_{1}$ CFT can be realized in a SO(5) spin model~\cite{Reshetikhin1983,Reshetikhin1985,TuHH2008,Alet2011,TuHH2011}, but it is more convenient to study the spin-$2$ Hamiltonian
\begin{eqnarray}
    H_{S=2} = \frac{25}{18} \sum_{j} \left[ -\mb{S}_{j} \cdot \mb{S}_{j+1} -\frac{1}{30} \left( \mb{S}_{j} \cdot \mb{S}_{j+1} \right)^{2}  + \frac{1}{15} \left( \mb{S}_{j} \cdot \mb{S}_{j+1} \right)^{3} + \frac{1}{150}\left( \mb{S}_{j} \cdot \mb{S}_{j+1} \right)^{4}\right],
\end{eqnarray}
which has SO(5) symmetry if spin-2 is regarded as the vector representation of SO(5). This model is the critical point between a SO(5) SPT phase and a trivial dimerized phase~\cite{TuHH2008}. It is also possible to design two ways to explore the screening physics. However, if a SPT state is used to generate impurities, the computational cost for PBC would be too high. Firstly, it is well-known that DMRG is more difficult for periodic systems. Secondly, the bond dimension of the matrix product operator representing the Hamiltonian also increases considerably when using PBC. Due to these complications, we have only studied the systems as shown in Fig.~1 (a) of the main text. The edge coupling is also described by Eq.~\eqref{eq:EdgeCouple} but now $\mb{S}_{1,2}$ are spin-2 operators. According to the identifications in Eq.~\eqref{Spin5TDL}, the boundary state $\ket{{\cal D}_1}$ may be realized. Below we present numerical evidence in favor of this conjecture.

\begin{figure}[ht]
\centering
\includegraphics[width=0.85\textwidth]{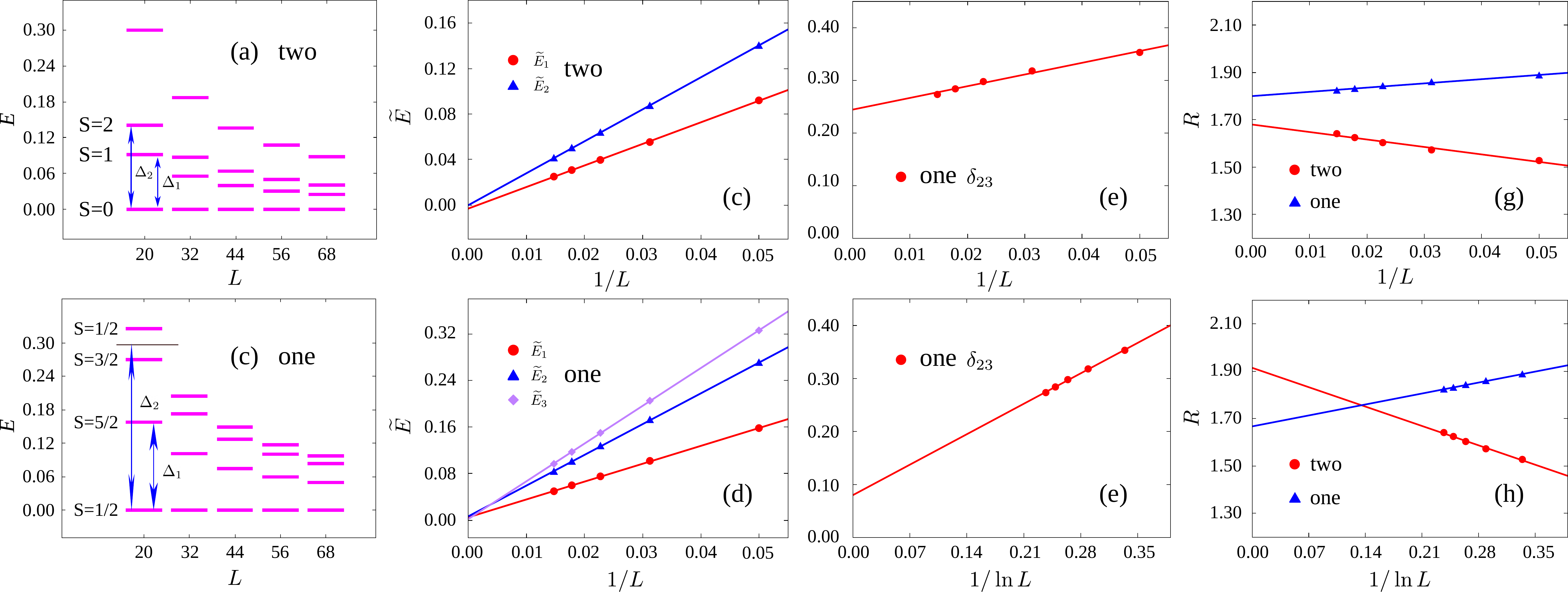}
\caption{Numerical results on the spin-2 Reshetikhin chains with spin-1/2 impurities. (a,b) Energy spectra for the cases with two and one impurity. The symbols are defined similarly as in Fig.~\ref{fig:ULSAKLT}. (c,d) Finite-size scaling analysis of the energy spacings $\widetilde{E}_{i}=E_{i}-E_{0}$. (e,f) Finite-size scaling analysis of the energy splitting $\delta_{23}$ for one impurity. (g,h) Finite-size scaling analyis of the ratios $R=\Delta_{2}/\Delta_{1}$.}
\label{fig:Reshetikhin}
\end{figure}

\begin{table}[ht]
\begin{tabular}{c|ccccc}
\hline
\hline
$L$         & 20 & 32 & 44 & 56 & 68 \\
            \cline{2-4}          
\hline
two impurites            &  1.527  &  1.574  &  1.603  &  1.625  &  1.642 \\
one impurity            &  1.887  &  1.861  &  1.844  &  1.832  &  1.823 \\
\hline
\hline
\end{tabular}
\caption{Numerical values of $R=\Delta_{2}/\Delta_{1}$ in the spin-2 Reshetikhin chains with spin-1/2 impurities.}
\label{TableS2}
\end{table}

Numerical results for $J_{1}=2$ and $J_{2}=-0.2$ are displayed in Fig.~\ref{fig:Reshetikhin}. The number of sites $L$ should be even to ensure that trivial boundary conditions are realized in the absence of impurities. When both ends are attached with spin-1/2, the first three levels have $S=0,1,2$, in good agreement with the BCFT prediction for ${\cal Z}_{{\cal D}_1, {\cal D}_1}$. If only one end is attached with spin-1/2, the first four levels have $S=\frac{1}{2},\frac{5}{2},\frac{1}{2},\frac{3}{2}$. To be consistent with the BCFT prediction for ${\cal Z}_{{\cal D}_1,\mathbf{0}}$, we need to assume that the last two levels collapse to a single level in the thermodynamic limit. As in the spin-1 examples, the energy spacings are inspected in Fig.~\ref{fig:Reshetikhin} (c,d), which exhibit good linear decay with $L$. Since there is no quasi-degeneracy for the case with two impurities, we simply define $\Delta_{1}=E_{1}-E_{0},\Delta_{2}=E_{2}-E_{0}$. For the case with one impurity, $\delta_{23}=(E_{3}-E_{2})/\widetilde{E}_{1}$ is plotted in Fig.~\ref{fig:Reshetikhin} (g,h) to check if $E_{2}$ and $E_{3}$ indeed collapse. When $1/\ln{L}$ is used as the horizontal axis, the fitting quality is good but the extrapolated value is positive and not close to zero. This fact undermines our conjecture about their degeneracy and is tentatively attributed to strong finite-size effects. While $E_{2}$ and $E_{3}$ may not be quasi-degenerate based on existing results, we still proceed to study $\Delta_{1}=E_{1}-E_{0}$ and $\Delta_{2}=(E_{3}+E_{2})/2-E_{0}$. For both cases, the ratios $R=\Delta_{2}/\Delta_{1}$ are presented in Fig.~\ref{fig:Reshetikhin} (g,h) and also given in Table~\ref{TableS2}. The finite-size values deviate substantially from the expected numbers, and the extrapolated values do not improve much either. We have checked many different combinations of $J_{1}$ and $J_{2}$. For various $J_{1} \in [1,10]$ and $J_{2}=0$, the lowest three levels in the presence of two spin-1/2 impurities have $S=0,1,2$. The ratio at the same $L$ increases as $J_{1}$ gets smaller but is never close to $3$ for the available sizes. If there is one impurity, the same parameters may not give the expected quantum numbers for small $L$. For $J_{1}=2$, the fourth eigenstate has $S=7/2$ at $L=12$ but changes to $S=1/2$ at $L=16$. If $J_{1}$ increases, a larger $L$ is needed to ensure that the fourth eigenstate has $S=1/2$. If we turn on a small negative $J_{2}$, finite-size effects are suppressed, but the ratios are still quite different from the theoretical predictions.

These findings, particularly the large splittings and significant deviations of $R$ from the expected values, clearly indicate very strong finite-size effects. One possible source for the discrepancy is the marginally irrelevant terms in the low-energy theory, which is common in the presence of continuous symmetry. Physical quantities are modified by logarithmic corrections such that they flow very slowly toward the values predicted by CFT. Another possibility is the competition between the $\cal{D}_1$ and $\sigma$ boundary conditions, both of which can screen half-integer spin impurity. In fact, the $\sigma$ boundary condition has a lower Affleck-Ludwig entropy, so presumably it is more stable in the renormalization group (RG) sense. However, $\ket{\sigma}$ is a Cardy state for Spin(5)$_1$ with a higher emergent symmetry than $\ket{\cal{D}_1}$, so it may require a longer RG flow. If this is indeed true, then what we have observed may be crossover phenomena. For the spin-1 ULS chain, there is no such complication since there is a unique spin-1/2 boundary state. In total, our results are consistent with the formation of an exotic conformal boundary condition described by the TDL $\cal{D}_{1}$ when a spin-1/2 impurity is screened.

\end{document}